%% file: main.tex
\documentclass[12pt]{l4dc2022}

\usepackage{times}
\usepackage[utf8]{inputenc}

\usepackage{algorithm}
\usepackage{algorithmic}

\usepackage{pgfplots}
\DeclareUnicodeCharacter{2212}{−}
\usepgfplotslibrary{groupplots,dateplot}
\usetikzlibrary{patterns,shapes.arrows}
\pgfplotsset{compat=newest}
\usepackage{hyperref}

\usepackage{hyperref}
\usepackage{url}

\usepackage{times}
\usepackage{wrapfig}

\usepackage{times}

\usepackage{booktabs}

\usepackage{mathrsfs} 
\usepackage{amsfonts,float}
\usepackage{mathtools}

\usepackage{color} 
\newtheorem{theorem1}{Theorem}

\newtheorem{assumption}{Assumption}
\newtheorem{sideinfo}{Side information}
\newtheorem{problem}{Problem}
\newtheorem{remark1}{Remark}

\usepackage{tikz,pgfplots}
\pgfplotsset{compat=1.8}
\usepackage{pgfplotstable}
\usepgfplotslibrary{groupplots}

\newcommand{\pU}{\textit{u}}

\newcommand{\Nn}{\mathbb{N}_{[1,n]}}
\newcommand{\Nd}{\mathbb{N}_{[1,d]}}
\newcommand{\lipfk}{\bar{L}_{f_k}}
\newcommand{\lipgpk}{\bar{L}_{g_{p,k}}}

\newcommand{\lipf}{\overline{f}}
\newcommand{\lipg}{\overline{g}}
\newcommand{\traj}{\mathscr{T}}

\newcommand{\dpg}{\texttt{D4PG}}
\newcommand{\sac}{\texttt{SAC}}
\newcommand{\sindyc}{\texttt{SINDYc}}

\usepackage[stable]{footmisc}
 
\title[Learning How to Reach, Swim, Walk and Fly in One Trial]{Learning How to Reach, Swim, Walk and Fly in One Trial: \\ Control of Unknown Systems with Scarce Data and Side Information}
\author{\Name{Franck Djeumou} \Email{fdjeumou@utexas.edu} \\
    \Name{Ufuk Topcu} \Email{utopcu@utexas.edu}\\
    \addr The University of Texas at Austin, United States
 }
 
\begin{document}

\maketitle

% MAX 9 CONTENT PAGES

\begin{abstract}
    We develop a learning-based control algorithm for unknown dynamical systems under very severe data limitations. Specifically, the algorithm has access to streaming and noisy data only from a single and ongoing trial. It accomplishes such performance by effectively leveraging various forms of side information on the dynamics to reduce the sample complexity. Such side information typically comes from elementary laws of physics and qualitative properties of the system. More precisely, the algorithm approximately solves an optimal control problem encoding the system's desired behavior. To this end, it constructs and iteratively refines a data-driven differential inclusion that contains the unknown vector field of the dynamics. The differential inclusion, used in an interval Taylor-based method, enables to over-approximate the set of states the system may reach. Theoretically, we establish a bound on the suboptimality of the approximate solution with respect to the optimal control with known dynamics. We show that the longer the trial or the more side information is available, the tighter the bound. Empirically, experiments in a high-fidelity F-16 aircraft simulator and MuJoCo's environments illustrate that, despite the scarcity of data, the algorithm can provide performance comparable to reinforcement learning algorithms trained over millions of environment interactions. Besides, we show that the algorithm outperforms existing techniques combining system identification and model predictive control.
    % We develop a learning-based control algorithm for unknown dynamical systems under very severe data limitations. Specifically, the algorithm has access to streaming and noisy data only from a single and ongoing trial. Despite the scarcity of data, we show---through a series of examples---that the algorithm can provide performance comparable to reinforcement learning algorithms trained over millions of environment interactions, while outperforming existing techniques combining system identification and model predictive control. It accomplishes such performance by effectively leveraging various forms of side information on the dynamics to reduce the sample complexity. Such side information typically comes from elementary laws of physics and qualitative properties of the system. More precisely, the algorithm approximately solves an optimal control problem encoding the system's desired behavior. To this end, it constructs and iteratively refines a data-driven differential inclusion that contains the unknown vector field of the dynamics. The differential inclusion, used in an interval Taylor-based method, enables to over-approximate the set of states the system may reach. Theoretically, we establish a bound on the suboptimality of the approximate solution with respect to the optimal control with known dynamics. We show that the longer the trial or the more side information is available, the tighter the bound. Empirically, experiments in a high-fidelity F-16 aircraft simulator and MuJoCo's environments illustrate the algorithm's effectiveness.%\looseness=-1
\end{abstract}


\input{tex/01_introduction}
\input{tex/02_background}
\input{tex/03_problem_statement}
\input{tex/04_reachable_set}
\input{tex/05_optimal_control}
\input{tex/06_numerical_experiments}
\input{tex/07_conclusion}

\bibliography{bibliography.bib}

\input{tex/08_appendix}

\end{document}

%% file: tex/01_introduction.tex
\section{Introduction}

Learning how to achieve a complex task has found numerous applications ranging from robotics~\citep{lillicrap2015continuous,schulman2015trust,deisenroth2013survey} to fluid dynamics~\citep{kutz2017deep}. However, learning algorithms generally suffer from high sample complexity, often requiring millions of samples to achieve the desired performance~\citep{nagabandi2018neural,schulman2015trust}. Such data requirements limit the practicability of learning algorithms in real-world scenarios where an excessive number of trials cannot be performed on a  physical system. 
A rather extreme example of such a scenario is an aircraft trying to retain a certain degree of control after abrupt changes in its dynamics, e.g., due to the loss of an engine. In such a scenario, there is a need to learn the dynamics after the abrupt changes using data from only the current trajectory.

We develop a learning-based control algorithm that utilizes data from a single trial and leverages side information on the unknown dynamics to reduce the sample complexity. The data include finitely many \emph{noisy} samples of the states, the states' derivatives, and the control signals applied. Under such a severe limitation on the amount of available data, learning can be performed efficiently only by incorporating already known invariant properties of the dynamical system. We refer to such extra knowledge as \emph{side information}. The side information, typically derived from elementary laws of physics, may be a priori knowledge of the regularity of the dynamics, monotonicity or bounds on the vector field, algebraic constraints on the states, or knowledge of parts of the vector field.

The developed algorithm, using the data and side information available to it, computes an over-approximation of the set of states the system may reach. Then, it incorporates such an over-approximation into a constrained short-horizon optimal control problem, which is solved on the fly.\looseness=-1
\begin{figure}[t]
    \centering
    \includegraphics[width=\textwidth,height=0.75in]{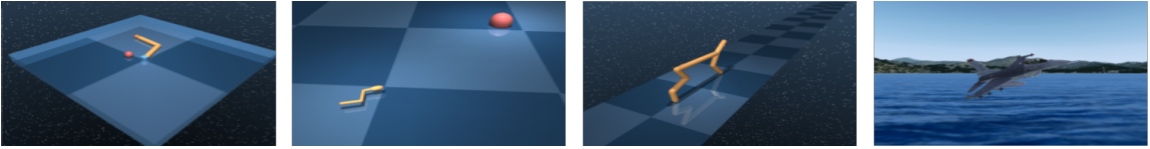}
    \vspace*{-1.2em}
    \caption{The developed learning-based algorithm can achieve near-optimal control of simulated robots and an F-16 aircraft using streaming data obtained from the systems' ongoing trajectory and side information derived from laws of physics. From left to right, we have the Reacher, Swimmer, Cheetah, and F-16 aircraft simulator environments.}
    \vspace*{-1.8em}
    \label{fig:environments}
\end{figure}

Specifically, it leverages a data-driven differential inclusion to compute over-approximations of the reachable sets of the system. It first constructs a differential inclusion that contains the unknown vector field. Next, it builds on set contractor programming~\citep{CHABERT20091079} to refine the differential inclusion as more data become available. Then, it computes over-approximations of the reachable sets of all dynamics described by the differential inclusion through an interval Taylor-based method~\citep{berz1998verified,nedialkov1999validated} that can enforce constraints from the side information to reduce the width of the over-approximations.

The obtained over-approximations enable to formulate the data-driven optimal control problem as a nonconvex and uncertain optimization problem. Specifically, we encode the control task as a sequential optimization of a cost function over a time horizon. Even for convex cost functions, the control problem is typically nonconvex. Besides, the predictions of the states' values at future times cannot be computed due to the unknown dynamics. The developed algorithm leverages the obtained over-approximations to optimize the nonconvex problem under the uncertain states' predictions.%\looseness=-1

The algorithm computes approximate solutions to the nonconvex optimization problem through convex relaxations. We develop a sequential convex optimization scheme~\citep{mao2019successive} that uses the obtained over-approximation and iteratively linearizes its nonconvex constraint around the previous iteration solution. Thus, each iteration solves a \emph{convex} optimization problem, and we leverage trust regions to account for the potential errors due to the linearization.

Theoretically, we establish a bound on the suboptimality of the approximate solution with respect to the optimal control solution in the case where the dynamics were known. The bound is proportional to the width of the obtained over-approximations. We show that the longer the trial or the more the side information available, the tighter the over-approximations. Thus, the algorithm achieves near-optimal control as more data streams or more side information is available.

Empirically, through a series of simulation examples, we show that the algorithm can provide performance comparable to reinforcement learning (RL) algorithms, such as \dpg{}~\citep{barthmaron2018distributed} and \sac{}~\citep{haarnoja2018soft}, while outperforming the system identification technique with model predictive control \sindyc{} \citep{brunton2016sparse,kaiser2018sparse}. We train \sac{} and \dpg{} over millions of environment interactions before comparing to our approach. \emph{We emphasize that if we had made fair comparisons, i.e., the baselines RL algorithms were also trained using streaming data from only the ongoing and single episode, the developed algorithm would have achieved significantly higher performance than any of these baselines since they cannot learn with such constraints on the amount of data}. Specifically, in several control tasks from MuJoCo~\citep{todorov2012mujoco,tassa2018deepmind}, we provide promising and comparative results to \dpg{} and \sac{}. Further, in a ground collision avoidance scenario of an F-16 aircraft~\citep{heidlauf2018verification}, we show that the algorithm outperforms \sindyc{} and the tuned F-16's linear-quadratic regulator controller.\looseness=-1

\paragraph{Related Work.}
In our prior work~\citep{djeumou2021fly, djeumou2020onthefly}, we described a data-driven algorithm similar to the algorithm developed in this paper. However, the algorithm~\citep{djeumou2021fly, djeumou2020onthefly} works only for control-affine dynamics. Further, most of the considered side information is not tailored for robotics systems, and only one-step optimal control problems were investigated. In contrast, the algorithm in this paper is applicable for a more general class of dynamics with polynomial dependency in control. We also evaluate the developed algorithm on highly-complex systems and 
consider a larger set of side information, e.g., algebraic constraints on states and unknown terms. Besides, we investigate short-horizon rather than one-step optimal control problems.

Several approaches for data-driven control combine model predictive control with system identification or data-driven reachable set estimation. These approaches achieve system identification through sparse regression over a library of nonlinear functions~\citep{kaiser2018sparse}, regression over the set of polynomials of fixed degree with physics-based side information~\citep{ahmadi2020learning}, spectral properties of the collected data~\citep{proctor2016dynamic}, Koopman theory~\citep{korda2018linear}, or Gaussian processes~\citep{krause2011contextual,pmlr-v120-gahlawat20a}. The approaches~\citep{devonport2020data,haesaert2017data,chakrabarty2018data} achieve data-driven estimation of the reachable sets of partially unknown dynamics using either supervised learning or Gaussian processes. They provide only probabilistic guarantees of the correctness of the computed reachable sets while \emph{our algorithm computes correct over-approximations}. Recent work~\citep{berberich2020data,berberich2020combining,markovsky2021data,van2020noisy,van2020data} and \texttt{DeePC}~\citep{Coulson2019DataEnabledPC}  have proposed data-driven control techniques based on the behavioral systems theory foundation~\citep{1428856}, which bypass the system identification step. These techniques mostly assume linear time-invariant dynamical systems and are extremely performant in such a setting. \emph{Except for \cite{ahmadi2020learning} that considers limited  side information and builds on computationally expensive semidefinite programs solvers, none of the above approaches (in their current form) can exploit the side information in this paper. Besides, through extensive comparisons with \sindyc{}, \texttt{DeePC}, and Gaussian-based approaches,~\cite{djeumou2021fly, djeumou2020onthefly} empirically demonstrates that: (a) For simple systems such as a unicycle, these techniques achieve significant lower performance (computation time and control suboptimality) than an approach that can exploit side information; (b) These techniques struggle to learn on high-dimensional and complex systems (e.g., quadrotor). Thus, this paper compares against RL techniques even though they work in a drastically different regime of data.\looseness=-1
% Thus, we choose RL algorithms as the baselines due to such empirical results and the complexity of the systems in this paper.
}

Model-free~\citep{mnih2015human,oh2016control,lillicrap2015continuous,mnih2016asynchronous,schulman2015high} and model-based~\citep{nagabandi2018neural,deisenroth2011pilco, gu2016continuous, boedecker2014approximate, levine2014learning,ko2009gp} RL algorithms have been widely used for data-driven control of complex systems. Model-free algorithms can achieve high performance at the expense of high sample complexity~\citep{schulman2015high} while model-based algorithms are more data-efficient but are conservative and generally achieve lower performance than model free approaches. In contrast, our algorithm can work with data from only the system's current trajectory, and increases the data efficiency through side informationon the dynamics.

%% file: tex/02_background.tex
\section{Background}

\paragraph{Notation.} We denote an interval by $[a , b] = \{ x \in \mathbb{R} |  a \leq x \leq b \}$ for some $a, b\in \mathbb{R}$ such that $a \leq b$, the set $\{i,\hdots,j\}$ by $\mathbb{N}_{[i,j]}$ for $i,j \in \mathbb{N}$ with $i\leq j$, the $k^\mathrm{th}$ component of a vector $x$ and the $(k,j)$ component of a matrix $X$ by ${x}_k$ and ${X}_{k,j}$, respectively, the weighted norm of a vector $x \in \mathbb{R}^n$ by $||x||_w = \sqrt{\sum_{i=1}^n (w_i x_i)^2}$ for some $w \in \mathbb{R}^n_+$, and the Lipschitz constant of $f : \mathcal{X} \to \mathbb{R}$ by $L^w_f = \sup \{ L \in \mathbb{R} \: \vert \: |f(x) - f(y)| \leq L \|x-y\|_w, x,y \in \mathcal{X}, x \neq y\}$ for $\mathcal{X} \subseteq \mathbb{R}^n$.%\looseness=-1

\paragraph{Interval Analysis.}
We denote the set of intervals on $\mathbb{R}$ by $\mathbb{IR} = \{ \mathcal{A} = [\underline{\mathcal{A}},\overline{\mathcal{A}}]  \: | \: \underline{\mathcal{A}},\overline{\mathcal{A}} \in \mathbb{R}, \underline{\mathcal{A}} \leq \overline{\mathcal{A}}\}$, the set of $n$-dimensional interval vectors by $\mathbb{IR}^n$, and the set of $n\times m$-dimensional interval matrices by $\mathbb{IR}^{n \times m}$. We carry forward the definitions~\citep{moore1966interval} of arithmetic operations, set inclusion, and intersections of intervals to interval vectors and matrices by applying them componentwise. We use the term interval to specify an interval vector or interval matrix when it is clear from the context. 
Given $f : \mathcal{X} \mapsto \mathcal{Y}$ with $\mathcal{X} \subseteq \mathbb{R}^n$ and $\mathcal{Y} \subseteq \mathbb{R}^{m}$, we define an \emph{interval extension of $f$} as $\boldsymbol{f} : \mathbb{IR}^n \mapsto \mathbb{IR}^m$ satisfying $\boldsymbol{f}(\mathcal{A}) \supseteq \mathscr{R}(f, \mathcal{A}) = \{ f(x) \: | \: x \in \mathcal{A} \}, \: \forall \mathcal{A} \subseteq \mathcal{X}$.
Thus, given an interval $ \mathcal{A}$, $\boldsymbol{f}(\mathcal{A})$ is an interval that \emph{over-approximates} the range of values taken by $f$ over $\mathcal{A}$.

\paragraph{Interval-Based Contractor.}
Interval-based contractor programming is a mathematical framework to solve constraints involving interval variables. Given an initial over-estimation of the constraint's solutions, a contractor filters such variable domains, i.e., reduces the interval of each variable, without loss of solutions of the constraints. Consider the constraint $h(\cdot) \leq 0$. Assume that $\mathcal{A}=[\mathcal{A}_1,\hdots,\mathcal{A}_n] \in \mathbb{IR}^n$ is a set containing the solutions. Then, the contractor operator computes $C^h_{\mathcal{A}} = [C^h_{\mathcal{A}_1},\hdots, C^h_{\mathcal{A}_n}] \in \mathbb{IR}^n$ such that $C^h_{\mathcal{A}_i} \subseteq \mathcal{A}_i, \forall i \in \mathbb{N}_{[1,n]}$ and  $h(x) > 0$ for all $ x \in \mathcal{A} \setminus C^h_{\mathcal{A}}$.

Several polynomial-time algorithms~\citep{benhamou-contractor1999,van1997numerica,trombettoni} have been developed to compute contractors associated with a given constraint. For example, \texttt{HC4-Revise}~\citep{benhamou-contractor1999} is a linear-time algorithm that provides optimal contractors when each variable appears only once in the constraint. In the following, we use $C^h_{\mathcal{A}}$ to refer to the contracted interval resulting from any of these algorithms.\vspace{-3mm}

%% file: tex/03_problem_statement.tex
\section{Problem Formulation}

This paper considers nonlinear dynamics with polynomial dependency in the control inputs as\vspace{-0.15cm}
\begin{align} 
    \dot{x} = f(x) + \sum\nolimits_{p=1}^d  g_p(x) \pU[\alpha^p], \label{eq:polynomial-control}
\end{align}
where $d \in \mathbb{N}$, $\alpha^p \in \mathbb{N}^m$ is known, the state $x: \mathbb{R}_+ \mapsto \mathcal{X}$ is a continuous-time signal evolving in $\mathcal{X} \in \mathbb{IR}^n$, $\pU[\alpha^p] = u_1^{\alpha_1^p} \cdots u_m^{\alpha_m^p}$ is a monomial with variables from the control signal $u : \mathbb{R}^+ \mapsto \mathcal{U}$ where $\mathcal{U} \in \mathbb{IR}^m$. The vector-valued functions $f = [f_k] : \mathbb{R}^n \mapsto \mathbb{R}^n$ and $g_p = [g_{p,k}] : \mathbb{R}^n \mapsto \mathbb{R}^{n}$ are considered to be \emph{nonlinear and unknown}. Note that even if the dynamics are not in the class above, Taylor expansion provides a tight approximation of the dynamics that lies in such a class.

\begin{assumption}[\textsc{Lipschitz systems}] \label{ass:lipshitz-bounds}
     Given a set $\mathcal{A} \subseteq \mathbb{R}^n$, $f_k$ and $g_{p,k}$ admit local Lipschitz constants $L^w_{f_k},  L^{w}_{g_{p,k}}> 0$ on $\mathcal{A}$, for some $w \in \mathbb{R}^n_+$ and for all $k \in \Nn, p \in \Nd$.
\end{assumption}
Assumption~\ref{ass:lipshitz-bounds} is common in the framework of optimal control. We emphasize that even though we use the weighted norm to define the Lipschitz constants, the results of this paper can be straightforwardly extended to general modulus of continuity assumption on $f$ and $g_p$. The weighted norm has the advantage of providing information on the relative importance of each variable in the function. 

Besides, the domain $\mathcal{X} \in \mathbb{IR}^n$ is bounded. Thus, by Assumption~\ref{ass:lipshitz-bounds}, $f_k$ and $g_{p,k}$ admit global Lipschitz constants on $\mathcal{X}$. We exploit such a knowledge by assuming \emph{known upper bounds} on the Lipschitz constants. That is, we have access to $\lipfk \in \mathbb{R}_+$ and $\lipgpk \in \mathbb{R}_+$ as known upper bounds on the Lipschitz constants $L^w_{f_k}$ and $L^{w}_{g_{p,k}}$, respectively, for $k \in \Nn$ and $p \in \Nd$. We emphasize that the Lipschitz bounds can be directly estimated from data at the expense of weakening some of the guarantees in this paper. Our numerical experiments use Lipschitz bounds estimated from data.

In a discrete-time setting, we denote the initial time by $t_1 \geq 0$ and the current time by $t_j > t_1$ for some $j > 1$. Let $\traj_j = \{(\Tilde{x}^i,\Tilde{\dot{x}}^i, u^i)\}_{i=1}^{j-1}$ be the finite-length set of observations obtained between $t_1$ and $t_j$. The dataset $\traj_j$ contains $j-1$ \emph{noisy} samples of the exact state $x^i=x(t_i)$, the derivative $\dot{x}_i=\dot{x}(t_i)$ of the state, and the applied input $u^i = u(t_i)$. We build on the widely-used \emph{bounded} noise assumption and consider that $|x(t) - \Tilde{x}(t)| \leq \eta, \: |\dot{x}(t) - \Tilde{\dot{x}}(t)| \leq \bar{\eta}$
for all $t \in \mathbb{R}_+$ and for some vector values $\eta, \bar{\eta} \in \mathbb{R}_+^n$. Here the absolute value and the comparison are conducted elementwise.

We seek to control the unknown dynamical system~\eqref{eq:polynomial-control} by finding $u^j,\hdots, u^{j+N} \in \mathcal{U}$ that are solutions of the \emph{$N$-step optimal control problem}\vspace{-2mm}
\begin{align}
    \underset{u^j,\hdots,u^{j+N}\in \mathcal{U}}{\mathrm{minimize}}& \quad \sum\nolimits_{q=j}^{j+N} c(x^q, u^q, x^{q+1} = x(t_{q+1}; x_q, u_q)),\vspace{-2mm}\label{eq:n-step-optimal-control}
\end{align}
where $N$ is the planning horizon, $c$ is a known cost function, $x^j = x(t_j)$ is the known current state of the system, $t_q = t_j + (q-j) \Delta t $, $\Delta t$ is a constant time step, and $x^{q+1} = x(t_{q+1}; x^q, u^q)$ is the state at $t_{q+1}$, i.e., a solution of the differential equation ~\eqref{eq:polynomial-control} at $t_{q+1}$ when $x^q$ is the initial state and $u^q$ is the constant control applied between $[t_q,t_{q+1}]$. The optimization problem~\eqref{eq:n-step-optimal-control} is generally nonconvex since the state at $t_{q+1}$ is nonconvex due to the nonlinear dynamics. Besides, $x^{q+1}$ cannot be computed due to the unknown dynamics.
% As a consequence, we can only hope to approximate solutions of such an optimization problem.
\begin{problem}\label{prob:control}
	Given the dataset $\traj_{j}$, the current state $\Tilde{x}^j$, compute an approximate solution to the $N$-step optimal control problem~\eqref{eq:n-step-optimal-control} and characterize the suboptimality of such approximation.
\end{problem}

%% file: tex/04_reachable_set.tex
\section{Reachable Set Over-Approximation via Data-Based Differential Inclusions}
In this section, we first construct a differential inclusion $\dot{x} \in \boldsymbol{f}(x) + \sum_{p=1}^d \boldsymbol{g}_p(x) \pU[\alpha^p]$ that contains the unknown vector field. Then, we adapt an interval Taylor-based method to over-approximate the reachable set of dynamics described by the constructed differential inclusion. Finally, we show how additional side information constrains the Taylor expansion to provide tighter over-approximations.
% f(x^i) \in C_{\mathcal{F}^i}, g_p(x^i) \in C_{\mathcal{G}^i}
\begin{lemma}[\textsc{Over-approximation of $f$ and $g_p$}]\label{lem:overapprox-fg}
	Let the set $\mathscr{E}_j = \{ (\Tilde{x}^i,C_{\mathcal{F}^i}, C_{\mathcal{G}^i}) \}_{i=0}^{j-1}$ be such that $C_{\mathcal{F}^i} = [C_{\mathcal{F}_k^i}] \in \mathbb{IR}^n$ and $C_{\mathcal{G}^i} = [C_{\mathcal{G}_{p,k}^i}] \in \mathbb{IR}^{d \times n}$ satisfy $f_k(\Tilde{x}^i) \in C_{\mathcal{F}_k^i}$ and $g_{p,k}(\Tilde{x}^i) \in C_{\mathcal{G}_{p,k}^i}$ for all $p \in \Nd$ and $k \in \Nn$. Then, the interval-valued functions $\boldsymbol{f} = [\boldsymbol{f}_k] : \mathbb{IR}^n \to \mathbb{IR}^{n}$ and $\boldsymbol{g}_{p} = [\boldsymbol{g}_{p,k}] : \mathbb{IR}^n \to \mathbb{IR}^{n}$, defined by $\boldsymbol{f}_k(\mathcal{A}) =  \bigcap_{(\Tilde{x}^i, C_{\mathcal{F}^i}, \cdot) \in \mathscr{E}_j} C_{\mathcal{F}_k^i} + [-1,1] \lipf_k \boldsymbol{\eta}^w(\mathcal{A} - \Tilde{x}^i)$ and $\boldsymbol{g}_{p,k}(\mathcal{A}) = \bigcap_{(\Tilde{x}^i, \cdot, C_{\mathcal{G}^i}) \in \mathscr{E}_j} C_{\mathcal{G}_{p,k}^i} + [-1,1] \lipg_{p,k} \boldsymbol{\eta}^w(\mathcal{A} - \Tilde{x}^i)$, are such that $\mathscr{R}(f_k,\mathcal{A}) \subseteq \boldsymbol{f}_k(\mathcal{A})$ and $\mathscr{R}(g_{p,k}, \mathcal{A}) \subseteq \boldsymbol{g}_{p,k}(\mathcal{A})$ for all $\mathcal{A} \subseteq \mathcal{X}$. Furthermore, the function $\boldsymbol{\eta}^w : \mathbb{IR}^n \mapsto \mathbb{IR}$ can be any straightforward interval extension of the weighted norm $||\cdot||_w$.
% 	We have
% 	\begin{align}
% 	\boldsymbol{\eta}^w(\mathcal{S}) = \beta_1 \Big( \sum_{i=1}^n w_i \beta_2(\mathcal{S}_i) \Big), \: \forall \mathcal{S} \in  \mathbb{IR}^n,
% 	\end{align}
% 	where the functions $\boldsymbol{\beta}_1: \mathbb{IR} \mapsto \mathbb{IR}$ and $\boldsymbol{\beta}_2: \mathbb{IR} \mapsto \mathbb{IR}$ are interval extensions of $\sqrt{\cdot}$ and $(\cdot)^2$, respectively. For any $\mathcal{S} \in \mathbb{IR}$, we have that
% 	\begin{align}
% 	\boldsymbol{\beta}_1(\mathcal{S}) &= [\sqrt{\underline{\mathcal{S}}}, \sqrt{\overline{\mathcal{S}}}], \: \: \: \text{if } \underline{\mathcal{S}} \geq 0 , \label{eq:sqrt_ext}\\
% 	\boldsymbol{\beta}_2(\mathcal{S}) &= \begin{cases} [0 , \max\{\underline{\mathcal{S}}^2,\overline{\mathcal{S}}^2\}],& \text{if } 0 \in \mathcal{S} \\
% 	[\min \{\underline{\mathcal{S}}^2,\overline{\mathcal{S}}^2\}, \max\{\underline{\mathcal{S}}^2,\overline{\mathcal{S}}^2\}],& \text{otherwise}.\end{cases}\label{eq:sqr_ext}
% 	\end{align}
\end{lemma}

We provide a proof of the lemma and an expression for $\boldsymbol{\eta}^w$ in the extended version of the paper \citep{djeumou2021learning}. Intuitively, Lemma~\ref{lem:overapprox-fg} states that if a set $\mathscr{E}_j = \{ (\Tilde{x}^i,C_{\mathcal{F}^i}, C_{\mathcal{G}^i}) \}_{i=0}^{j-1}$ is known, it is possible to obtain an analytic formula to over-approximate the unknown $f$ and $g_p$ via the Lipschitz bounds. Lemma~\ref{lem:contraction} enables to compute the set $\mathscr{E}_j$ based on the data $\mathscr{T}_j$.\vspace{-3mm}

\begin{lemma}[\textsc{Refinement via contractor}] \label{lem:contraction}
    Given a data point $(\Tilde{x}^i,\Tilde{\dot{x}}^i, u^i) \in \traj_j$, an interval $\mathcal{F}^i = [\mathcal{F}^i_k] \in \mathbb{IR}^n$ such that $f(\Tilde{x}^i) \in \mathcal{F}^i$, and an interval $\mathcal{G}^i = [\mathcal{G}^i_{p,k}] \in \mathbb{IR}^{d \times n}$ such that $g_{p,k}(\Tilde{x}^i) \in \mathcal{G}_{p,k}^i$ for all $p \in \Nd,k \in \Nn$. Let the intervals $C_{\mathcal{F}^i} \in \mathbb{IR}^n$ and $C_{\mathcal{G}^i} \in \mathbb{IR}^{d\times n}$ defined by
	\small
    \begin{align*}
        C_{\mathcal{F}^i_k} &= \mathcal{F}^i_k \: \cap \: \left\{\Tilde{\mathcal{N}}^i_k - \sum_{p=1}^d \mathcal{G}^i_{p,k} \pU^i[\alpha^p] \right\}, C_{\mathcal{G}^i_{p,k}} = \begin{cases} \Big( \left\{ \mathcal{S}_{p-1,k} - \sum_{l = p+1}^d \mathcal{G}^i_{l,k} \pU^i[\alpha^l] \right\} \cap   \mathcal{G}^i_{p,k} \pU^i[\alpha^p] \Big) \frac{1}{\pU^i[\alpha^p]}, \\ 
        \quad \mathrm{if} \ \pU^i[\alpha^p] \neq 0, \\
        \mathcal{G}^i_{p,k}, \quad \mathrm{otherwise},
        \end{cases}\\
        \mathcal{S}_{0,k} &= \left\{ \Tilde{\mathcal{N}}^i_k - C_{\mathcal{F}^i_k} \right\} \cap \left\{ \sum_{p=1}^d \mathcal{G}^i_{p,k} \pU^i[\alpha^p] \right\},
        \mathcal{S}_{p,k} = \left\{ \mathcal{S}_{p-1,k} - C_{\mathcal{G}^i_{p,k}} \pU^i[\alpha^p] \right\} \cap \left\{ \sum_{l = p+1}^d \mathcal{G}^i_{l,k} \pU^i[\alpha^l] \right\},
    \end{align*}
    % \begin{subequations} \label{eq:contraction-fG}
    %     \begin{align}
    %         C_{\mathcal{F}^i_k} & = \mathcal{F}^i_k \: \cap \: \left\{\dot{x}^i_k - \sum_{p=1}^d \mathcal{G}^i_{p,k} \pU^i[\alpha^p] \right\}, \\
    %         \mathcal{S}_{0,k} & = \left\{ \dot{x}^i_k - C_{\mathcal{F}^i_k} \right\} \cap \left\{ \sum_{p=1}^d \mathcal{G}^i_{p,k} \pU^i[\alpha^p] \right\}, \\
    %         C_{\mathcal{G}^i_{p,k}} & = \begin{cases} \Big( \left\{ \mathcal{S}_{p-1,k} - \sum_{l = p+1}^d \mathcal{G}^i_{l,k} \pU^i[\alpha^l] \right\} \cap   \mathcal{G}^i_{p,k} \pU^i[\alpha^p] \Big) \frac{1}{\pU^i[\alpha^p]}, \\ 
    %         \quad \mathrm{if} \ \pU^i[\alpha^p] \neq 0, \\
    %         \mathcal{G}^i_{p,k}, \quad \mathrm{otherwise},
    %         \end{cases}\\
    %         \mathcal{S}_{p,k} & = \left\{ \mathcal{S}_{p-1,k} - C_{\mathcal{G}^i_{p,k}} \pU^i[\alpha^p] \right\} \cap \left\{ \sum_{l = p+1}^d \mathcal{G}^i_{l,k} \pU^i[\alpha^l] \right\}.
    %     \end{align}
    % \end{subequations}
    \normalsize
    for successive values of $k \in \Nn$ and for all $p \in \Nd$ with $\Tilde{\mathcal{N}}^i_k = [\Tilde{\dot{x}}_i - \bar{\eta},\Tilde{\dot{x}}^i + \bar{\eta}]$. Then, $C_{\mathcal{F}^i}$ and $C_{\mathcal{G}^i}$ are the smallest intervals enclosing $f(\Tilde{x}^i)$ and $g_p(\Tilde{x}^i)$, given only the data $(\Tilde{x}^i,\Tilde{\dot{x}}^i,u^i)$, $\mathcal{F}^i$, $\mathcal{G}^i$.\vspace{-8mm}
\end{lemma}
% \vspace{-0.20cm}
\begin{minipage}[t]{0.49\textwidth}
\begin{algorithm}[H]
    % \centering
    \caption{\texttt{Construct}: Compute $\mathscr{E}_j$ required to over-approximate $f$ and $g_p$ at each data point of a given trajectory.}
    \label{algo:construction}
    \textbf{Input}: Dataset $\traj_j$ and a parameter $M >0$. \\
    \textbf{Output}:${\;\mathscr{E}_j = \{ (\Tilde{x}^i,C_{\mathcal{F}^i}, C_{\mathcal{G}^i})\}_{i=0}^{j-1}}$.
    
    \begin{algorithmic}[1]  
    
        \STATE $\mathcal{A} \gets \mathcal{X}$, $\mathcal{R}^{f_\mathcal{A}}, \mathcal{R}^{G_\mathcal{A}}\gets [-M,M]^n$
        \STATE Define $\Tilde{x}^0 \in \mathcal{A}$, $C_{\mathcal{F}^0} \gets \mathcal{R}^{f_\mathcal{A}}$, $C_{\mathcal{G}^0} \gets \mathcal{R}^{G_\mathcal{A}}$ \label{eq:sideinfo-boundvectorfield}
        \FOR{$i \in \mathbb{N}_{[1,j]} \wedge (\Tilde{x}^i, \Tilde{\dot{x}}^i, u^i) \in \mathscr{T}_{j}$} 
            \STATE $\mathscr{E}_i \gets \texttt{Refine}((\Tilde{x}^i, \Tilde{\dot{x}}^i, u^i),\mathscr{E}_{i-1}, \traj_i)$
        \ENDFOR
    \STATE \Return $\mathscr{E}_j$
    
  \end{algorithmic}
  
\end{algorithm}
\end{minipage}
\hfill
\begin{minipage}[t]{0.49\textwidth}

\begin{algorithm}[H]
    \caption{\texttt{Refine}: Update $\mathscr{E}_j$ with data.}
    \label{algo:refine}
    \textbf{Input}: A point $(\Tilde{x}^j, \Tilde{\dot{x}}^j, u^j)$, $\mathscr{E}_j$ containing past over-approximations,  $\traj_j$, and the noise bound $\bar{\eta}$. \\
    \textbf{Output}:${\;\mathscr{E}_{j+1} = \{ (\Tilde{x}^i,C_{\mathcal{F}^i}, C_{\mathcal{G}^i})\}_{i=0}^{j}}$.
    
    \begin{algorithmic}[1]    
    
        \STATE Compute $\mathcal{F}^j = \boldsymbol{f}(\Tilde{x}^j), \mathcal{G}^j = [\boldsymbol{g}_{p,k}(\Tilde{x}^j)]$ via Lemma~\ref{lem:overapprox-fg} and $\mathscr{E}_{j}$ \label{alg:update-ei}
        \STATE Compute $C_{\mathcal{F}^{j}}, C_{\mathcal{G}^j}$ via Lemma~\ref{lem:contraction} \label{alg:contraction-f-G}
        % \WHILE{$\mathscr{E}_{j+1}$ is not invariant}
            \FOR{$(\Tilde{x}^i, \Tilde{\dot{x}}^i, u^i) \in \mathscr{T}_{j+1}$}
                \STATE Execute~\ref{alg:update-ei}--\ref{alg:contraction-f-G} with $j=i$, $\mathscr{E}_{j}= \mathscr{E}_{j+1}$ 
            \ENDFOR
        % \ENDWHILE 
    \STATE \Return $\mathscr{E}_{j+1}$
    
  \end{algorithmic}
  
\end{algorithm}
\end{minipage}

The proof of the lemma is provided in the extended version of the paper \citep{djeumou2021learning}. Lemma~\ref{lem:contraction} provides tighter sets $C_{\mathcal{F}^i} \subseteq \mathcal{F}^i$ and $C_{\mathcal{G}^i} \subseteq \mathcal{G}^i$ that prune out from $\mathcal{F}^i$ and $\mathcal{G}^i$ some values $f(\Tilde{x}^i)$ and $g_p(\Tilde{x}^i)$ that do not satisfy the dynamics constraint $\dot{x}^i = f(x^i) + \sum_{p=1}^d g_p(x^i) \pU^i[\alpha^p]$.\vspace{-2mm} %Next, we develop Algorithm~\ref{algo:construction} that utilizes $\traj_j$ to construct and refine the set $\mathscr{E}_j$ required in Lemma~\ref{lem:overapprox-fg} to over-approximate $f$ and $g_p$. Then, Theorem~\ref{thm:diffinclusion} constructs the desired data-based differential inclusion.%\looseness=-1
\begin{theorem1}[\textsc{Data-driven differential inclusion}] \label{thm:diffinclusion}
    Given a dataset $\traj_j$, the bounds $\lipf_k$ and $\lipg_{p,k}$, it holds that the unknown vector field of the dynamics~\eqref{eq:polynomial-control} satisfies\vspace{-0.2cm}
    \begin{align}
        \dot{x} \in \boldsymbol{h}(x,u) \triangleq \boldsymbol{f}(x) + \sum\nolimits_{p=1}^d \boldsymbol{g}_p(x) \pU[\alpha^p], \label{eq:diffinclusion}
    \end{align}\vspace{-0.05cm}
    where $\boldsymbol{f}$ and $\boldsymbol{g}_p$ are obtained from Lemma~\ref{lem:overapprox-fg} with $\mathscr{E}_j$ taken as the output of Algorithm~\ref{algo:construction}.\vspace{-2mm}
\end{theorem1}

% \begin{proof}
%     This result is straightforward from Lemma~\ref{lem:overapprox-fg} and Lemma~\ref{lem:contraction}. First, let $i \in \mathbb{N}_{[1,j]}$. We show that for all $(x^i, C_{\mathcal{F}^i}, C_{\mathcal{G}^i}) \in \mathscr{E}_j$ given by Algorithm~\ref{algo:construction}, we have $f_k(x^i) \in C_{\mathcal{F}_k^i}$ and $g_{p,k}(x^i) \in C_{\mathcal{G}_i}$ for all $p,k\in \Nd \times \Nn$ . Specifically, as a consequence of line~\ref{alg:update-ei} of Algorithm~\ref{algo:refine} and Lemma~\ref{lem:overapprox-fg}, we have that $f_k(x^i) \in \mathcal{F}_k^i$ and $g_{p,k}(x^i) \in \mathcal{G}^i_{p,k}$. Hence, by line~\ref{alg:contraction-f-G} of Algorithm~\ref{algo:refine} and Lemma~\ref{lem:contraction}, we immediately have that $f_k(x^i) \in C_{\mathcal{F}_k^i}$ and $g_{p,k}(x^i) \in C_{\mathcal{G}_{k,p}^i}$. Thus, $\mathscr{E}_j$ can be used in Lemma~\ref{lem:overapprox-fg} to conclude that $f_k(x) \in \boldsymbol{f}_k(x)$ and $g_{p,k}(x) \in \boldsymbol{g}_{p,k}(x)$ for all $x \in \mathcal{X}$. Therefore, we have $\dot{x} = f(x) + \sum_{p=1}^d g_p(x) \pU[\alpha^p] \in \boldsymbol{f}(x) + \sum_{p=1}^d \boldsymbol{g}_p(x) \pU[\alpha^p]$ through interval arithmetic.
% \end{proof} 
\begin{remark1}[\textsc{Persistent excitation}]
    The quality of the differential inclusion~\eqref{eq:diffinclusion} depends on how much information on $f$ and $g_p$ can be obtained from the dataset $\mathscr{T}_j$. This is the classical observability problem, sometimes referred to as persistent excitation. Thus, the learning algorithm should sometimes take suboptimal actions through persistent excitations of the system.
\end{remark1}    

Finally, we compute over-approximations of the reachable sets of all dynamics described by the differential inclusion~\eqref{eq:diffinclusion}. Theorem~\ref{thm:one-step-overapprox} provides a closed-form expression for such a set.
\begin{theorem1}[\textsc{Data-driven reachable set over-approximation}]\label{thm:one-step-overapprox}
    Given the dataset $\traj_j$, a constant control signal $u : t \mapsto u^q$ on the interval $[t_q,t_{q+1}]$ with $u^q \in \mathcal{U}$, and the uncertain set $\mathcal{R}^q \in \mathbb{IR}^n$ of states $x^q$ at time $t_q$. Then, a closed-form expression for $\mathcal{R}^{q+1} \supseteq \{x(t_{q+1}; u^q, x^q) \in \mathcal{X} | x^q \in \mathcal{R}^q\}$, which over-approximates the reachable set at $t_{q+1}$ for all $x^q \in \mathcal{R}^q$, is given by\vspace{-3mm}
    \begin{align}
        \mathcal{R}^{q+1} = \mathcal{R}^q +  \boldsymbol{h}(\mathcal{R}^q,u^q) \Delta t
        + \big( \mathcal{J}^f + \sum\nolimits_{p=1}^d \mathcal{J}^{g_p} \pU^q[\alpha^p] \big) \boldsymbol{h}(\mathcal{P}^q, u^q) \Delta t^2/2,\vspace{-2mm} \label{eq:nextstate}
    \end{align}
    where the matrices $\mathcal{J}^f = [\mathcal{J}_{k,l}^f] \in \mathbb{IR}^{n \times n}$ and $\mathcal{J}^{g_p} = [\mathcal{J}_{k,l}^{g_p}] \in \mathbb{IR}^{n \times n}$, over-approximations of the Jacobian of $f$ and $g_p$ , are such that $\mathcal{J}_{k,l}^f = [-1,1] w_k \lipf_k $ and $\mathcal{J}_{k,l}^{g_p} = [-1,1] w_k \lipg_{p,k} $ for all $p\in \Nd$ and $k,l \in \Nn$. Further, the set $\mathcal{P}^q$, a rough enclosure of $\{x(t_{q+1}; u^q, x^q) \in \mathcal{X} | x^q \in \mathcal{R}^q\}$, is a solution of the fixpoint equation $\mathcal{R}^{q} \ + \  [0, \Delta t]\ \boldsymbol{h}(\mathcal{P}^q, u^q) \ \subseteq \ \mathcal{P}^{q}$. 
    % \begin{align}
    %     \mathcal{R}^{q} \ + \  [0, \Delta t]\ \boldsymbol{h}(\mathcal{P}^q, u^q) \ \subseteq \ \mathcal{P}^{q}. \label{eq:fix-point-rough}
    % \end{align}
\end{theorem1}\vspace{-0.1cm}

We provide the proof of Theorem~\ref{thm:one-step-overapprox} in the extended paper~\citep{djeumou2021fly}. It merges the differential inclusion with an interval Taylor-based expansion of order $2$ to obtain the result.
% \begin{remark1}
%     Theorem~\ref{thm:one-step-overapprox} does not consider uncertain control inputs or \emph{noises} as in the literature of reachable sets' computation. However, such an extension is straightforward from the proof of the theorem. We show in the supplementary material how to incorporate \emph{noises} in this framework.%\looseness=-1
% \end{remark1}\vspace{-0.1cm}

Theorem~\ref{thm:one-step-overapprox}, as it is, does not incorporate side information other the regularity assumption. We describe in the following how to incorporate a-priori knowledge to tighten $\mathcal{R}^{q+1}$ given by~\eqref{eq:nextstate}.
\begin{sideinfo}[\textsc{Partial dynamics knowledge}]\label{side:partknowledge}
    The vector field of~\eqref{eq:polynomial-control} contains both known and unknown terms. That is, $\dot{x} = \sum_{s=1}^S \mathrm{f}^s(x) \cdot f^s(x) + \sum_{p=1}^d \sum_{s=1}^S \mathrm{g}^s_p(x) \cdot g_p^s(x) \pU[\alpha^p]$, where $\cdot$ denotes the elementwise product between vectors or matrices, $\mathrm{f}^s$ and $\mathrm{g}^s_p(x)$ are known functions, and $f^s, g_p^s$ are unknown functions satisfying Assumption~\ref{ass:lipshitz-bounds}.
    
    Given $\mathscr{E}^s_j$ containing past over-approximations of $f^s, g^s_p$ and a new data point $(\Tilde{x}^j,\Tilde{\dot{x}}^j, u^j)$, the refinement (Algorithm~\ref{algo:refine}) is adapted to compute in line~\ref{alg:update-ei} over-approximations $\boldsymbol{f}^s(\Tilde{x}^j)$ and $\boldsymbol{g}^s_p(\Tilde{x}^j)$ via Lemma~\ref{lem:overapprox-fg} and $\mathscr{E}^s_j$. Then, line~\ref{alg:contraction-f-G} is modified such that each $\boldsymbol{f}^s(\Tilde{x}^j)$ and $\boldsymbol{g}^s_p(\Tilde{x}^j)$ are contracted according to the new dynamics' constraint $\dot{x}^j = \sum_{s=1}^S \mathrm{f}^s(x^j) \cdot f^s(x^j) + \sum_{p=1}^d \sum_{s=1}^S \mathrm{g}^s_p(x^j) \cdot g_p^s(x^j) \textit{u}^j[\alpha^p]$. The contracted sets can be obtained straightforwardly by slight changes in the scheme described by Lemma~\ref{lem:contraction} or by calling an algorithm such as \texttt{HC4-Revise}. Thus, the new differential inclusion~\eqref{eq:diffinclusion} is given by $\boldsymbol{h}(x,u) = \sum_{s=1}^S \boldsymbol{\mathrm{f}}^s(x) \cdot \boldsymbol{f}^s(x) + \sum_{p=1}^d \sum_{s=1}^S \boldsymbol{\mathrm{g}}^s_p(x) \cdot \boldsymbol{g}_p^s(x) \pU[\alpha^p]$, where $\boldsymbol{\mathrm{f}}^s$ and $\boldsymbol{\mathrm{g}}^s$ are interval extensions of known $\mathrm{f}^s$ and $\mathrm{g}^s_p$.
    Furthermore, we compute the new Jacobian terms $\mathcal{J}^f$ and $\mathcal{J}^{g_p}$ used in $\mathcal{R}^{q+1}$ by applying chain rules and exploiting the Lipschitz bounds.
    % as follows: $\mathcal{J}^f = \sum_{s=1}^S \big( \frac{\partial \boldsymbol{\mathrm{f}}^s}{\boldsymbol{\partial x}}(\mathcal{P}^q) \cdot \boldsymbol{f}^s(\mathcal{P}^q) + \mathcal{J}^{f^s} \cdot \boldsymbol{\mathrm{f}}^s(\mathcal{P}^q) \big)$ and $\mathcal{J}^{g^p} = \sum_{s=1}^S \big( \frac{\partial \boldsymbol{\mathrm{g}}_p^s}{\boldsymbol{\partial x}}(\mathcal{P}^q) \cdot \boldsymbol{g}_p^s(\mathcal{P}^q) + \mathcal{J}^{g_p^s} \cdot \boldsymbol{\mathrm{g}}_p^s(\mathcal{P}^q) \big)$.%\vspace{0.10cm}
\end{sideinfo}

\begin{sideinfo}[\textsc{Algebraic constraints}]\label{sideinfo:algebraic}
    We are given a constraint $r(\dot{x}(\cdot), x(\cdot)) \geq 0$ where $r$ is a differential map. Such a constraint typically derives from conservation laws of physics. 
    
    Without loss of generality, we consider that $r : \mathbb{R}^n \times \mathbb{R}^n \mapsto \mathbb{R}$. This side information provides tighter over-approximations on $\boldsymbol{f}$, $\boldsymbol{g}_p$, $\mathcal{J}^f$, and $\mathcal{J}^{g_p}$ \emph{locally}. Specifically, this constraint can be formulated as the new constraint $w(f(x), [g_{p,k}(x)], u, x) \geq 0$. In some cases, another constraint $z(f(x), [g_{p,k}(x)], \frac{\partial f}{\partial x}(x), [\frac{\partial g_{p,k}}{\partial x}(x)], u, x) \geq 0$ can be derived by differentiating $w$. The new constraints $w$ and $z$ can be incorporated in the computation of $\mathcal{R}^{q+1}$ through contractors. More specifically, the refinement algorithm and the interval extensions of the Jacobian can be improved by additionally contracting with respect to the constraints $w$ and $z$. Thus, such side information enables to obtain a tighter $\mathcal{R}^{q+1}$. We develop on more side information in the extended paper.
\end{sideinfo}\vspace{-5mm}
% In general, any mathematical constraints on the states, its derivatives, or unknown terms of the dynamics can be enforced in the Taylor expansion using contractors. Due to the page limitations, we provide in the extended paper an ablation study on the side information.

%% file: tex/05_optimal_control.tex
\section{Approximate Optimal Control} \label{sec:optcontrol}

In this section, we develop an algorithm that computes approximate solutions to the optimal control problem~\eqref{eq:n-step-optimal-control} using over-approximations of the reachable sets. Further, we characterize the suboptimality of the approximate solutions with respect to the case of known dynamics.

The nonconvexity in the optimal control problem~\eqref{eq:n-step-optimal-control} is due to the possibly nonconvex cost function $c$ and the nonconvex constraint $x^{q+1} = x(t_{q+1}; x^q, u^q)$. We replace such an expression by $x^{q+1} = \hat{h}^{\theta}(x^q, u^q) \in \mathcal{R}^{q+1}$, where the function $\hat{h}^\theta$, parameterized with $\theta \in \mathbb{R}^n$, is a trajectory picked inside $\mathcal{R}^{q+1}$. For example, a straightforward choice can be $\hat{h}^\theta(x^q,u^q) = \theta \overline{\mathcal{R}^{q+1}} + (1-\theta) \underline{\mathcal{R}^{q+1}}$, for $\theta \in [0,1]^n$. Then, we solve the nonconvex problem by sequentially linearizing $x^{q+1}$ and the cost function $c$ around the solution of the $s^{\mathrm{th}}$ iteration. This results into a \emph{convex subproblem} that is solved to full optimality. The obtained solutions are then used at the $(s+1)^{\mathrm{th}}$ iteration.

\paragraph{Linearization.}
Let $\mathrm{x} = [x^{j+1};\hdots;x^{j+N+1}] \in \mathbb{R}^{nN}$ and  $\mathrm{u} = [u^{j};\hdots;u^{j+N}] \in \mathbb{R}^{mN}$. We denote the solutions of the $s^{\mathrm{th}}$ iteration by $\mathrm{x}^s = [x^{j+1,s};\hdots;x^{j+N+1,s}]$ and $\mathrm{u}^s = [u^{j,s};\hdots;x^{j+N,s}]$. Then, we can approximate the gradient of $h^\theta$ (or $x(t_{q+1}; x^q, u^q)$) around $\mathrm{x}^s, \mathrm{u}^s$ as follows:\vspace{-0.10cm}
\begin{align*}
    A^{q,s}  &= \frac{\partial h^\theta(x^q,u^q)}{\partial x^q} \Big\vert_{x^{q,s}, u^{q,s}}
        \in \mathbb{I} +  \big( \mathcal{J}^f(x^{q,s}) + \sum\nolimits_{p=1}^d \mathcal{J}^{g_p}(x^{q,s}) \textit{u}^{q,s}[\alpha^p]\big) \Delta t  = \mathcal{A}^{q,s},\\
    B^{q,s} &=  \frac{\partial h^\theta(x^q,u^q)}{\partial u^q} \Big\vert_{x^{q,s}, u^{q,s}}
        \in \sum\nolimits_{p=1}^d \Big[ \Delta t \boldsymbol{g}_{p,k}(x^{q,s})  \frac{\partial \pU[\alpha^p]}{\partial u_l}\Big\vert_{u^{q,s}} \Big]_{k,l} = \mathcal{B}^{q,s},
\end{align*}
where $\mathbb{I}$ is the identity matrix of appropriate dimensions. The jacobian $\mathcal{J}^f(x^{q,s}), \mathcal{J}^{g_p}(x^{q,s})$ are exactly $\mathcal{J}^f$ and $\mathcal{J}^{g_p}$ when no extra side information are given. With side information, the matrices are computed through chain rules as described in side information~\ref{side:partknowledge}. Note that since we neglect the term in $\Delta t^2$, $\mathcal{A}^{q,s}$ and $\mathcal{B}^{q,s}$ are approximations of the actual range of the gradients of $h^\theta$.

Next, we define the variables $\Delta \mathrm{x} = \mathrm{x} - \mathrm{x}^s$, $\Delta \mathrm{x}^{q} = x^q - x^{q,s}$, $\Delta \mathrm{u} = \mathrm{u} - \mathrm{u}^s$, and $\Delta \mathrm{u}^{q} = u^q - u^{q,s}$ in terms of the unknown solutions of the current iteration $\mathrm{x}$ and $\mathrm{u}$. Thus, at the $(s+1)^\mathrm{th}$ iteration, the first-order approximation of $x^{q+1} = \hat{h}^{\theta}(x^q, u^q)$ around the previous solution $(x^{q,s}, u^{q,s})$ is\vspace{-2mm}
\begin{equation}
    \vspace{-2mm}
    x^{q+1,s} + \Delta \mathrm{x}^{q+1} = \quad h^\theta (x^{q,s}, u^{q,s})  + A^{q,s} \Delta \mathrm{x}^q +B^{q,s} \Delta \mathrm{u}^q +  v^{q},\label{eq:linearization}
\end{equation}
where $\mathrm{v} = [v^{j};\hdots; v^{j+N}]$ are penalty variables that enable the linearization to be always feasible. Further, to ensure that the variable $v^q$ is used only when necessary, we augment the cost function with the sufficiently large penalization weight $\lambda > 0$. Thus, the solution for the $(s+1)^\mathrm{th}$ iteration, optimizes the penalized and linearized cost given by $L^s(\Delta \mathrm{x}, \Delta \mathrm{u}) = \sum\nolimits_{q=j}^{j+N} ( c(x^{q,s}, u^{q,s}, x^{q+1,s}) + \nabla c(x^{q,s}, u^{q,s}, x^{q+1,s}) [\Delta \mathrm{x}; \Delta \mathrm{u}]) + \lambda \sum\nolimits_{q=j}^{j+N} \|v^q\|,$
%  \begin{align}
%     %\small
%      L^s(\Delta \mathrm{x}, \Delta \mathrm{u}) = \sum\nolimits_{q=j}^{j+N} ( c(x^{q,s}, u^{q,s}, x^{q+1,s}) + \nabla c(x^{q,s}, u^{q,s}, x^{q+1,s}) [\Delta \mathrm{x}; \Delta \mathrm{u}]) + \lambda \sum\nolimits_{q=j}^{j+N} \|v^q\|,\vspace{-0.20cm} \nonumber
%  \end{align}
 where we also linearize the possibly nonconvex function $c$ given that $\nabla c$ is its gradient, and $\|\cdot\|$ can be either the infinity norm or $1$-norm. In order to verify the linearization accuracy, we also define the nonlinear \emph{realized} cost $J(\mathrm{x},\mathrm{u}) = \sum\nolimits_{q=j}^{j+N} c(x^{q}, u^{q}, x^{q+1}) + \lambda \sum\nolimits_{q=j}^{j+N} \|x^{q+1} - h^\theta(x^q, u^q) \|$.

\paragraph{Trust region constraints and linearized problem.} 
We impose the trust region constraint $\|\Delta \mathrm{u}\| \leq r^s$ to ensure that $\mathrm{u}$ does not deviate significantly from the control input $\mathrm{u}^s$ obtained in the previous iteration, where $r^s$ will be updated at each iteration so that the $\mathrm{x}$ remains close to $\mathrm{x}^s$. This update rule enables to keep the solutions within the region where the linearization is accurate. 
As a consequence, each iteration of our algorithm solves the following linear optimization problem:
 \begin{align}
     \underset{\Delta \mathrm{u}, \Delta \mathrm{x}}{\mathrm{minimize}}  &\quad L^s(\Delta \mathrm{x}, \Delta \mathrm{u}) \label{eq:convexsub}\\
     \mathrm{subject \ to} &\quad \eqref{eq:linearization}, \|\Delta \mathrm{u}\| \leq r^s, \mathrm{u}^s + \Delta \mathrm{u} \in \mathcal{U}^N, \mathrm{x}^s + \Delta \mathrm{x} \in \mathcal{X}^N.  \nonumber
 \end{align}
 The optimal solution of the linearized problem is either accepted and used in the next iteration or rejected until convergence. When the linearization is considered accurate, i.e., the realized cost $J$ and linearized cost $L^s$ are similar, the solution is accepted and the trust region is expanded. Otherwise, the solution is rejected and the trust region is contracted. 
\begin{theorem}[\textsc{Suboptimality bound}]\label{thm:suboptimality-gap}
    Assume that $L_c$ with the $2$-norm is the Lipschitz constant of the cost $c$ on $\mathcal{X} \times \mathcal{U} \times \mathcal{X}$. Let $C_j^{\star}$ and $\hat{C}_j$ be the optimal costs of the $N$-step control problem~\eqref{eq:n-step-optimal-control} when the dynamics are known, e.g. $x^{q+1} = x(\cdot, x^q, u^q)$ is known, and the dynamics are unknown, e.g., $x^{q+1} = h^\theta(x^q,u^q) \in \mathcal{R}^{q+1}$. Then, $|  C_j^* - C_j| \leq L_c \Big( \|\mathrm{wd}(\mathcal{R}_\mathcal{U}^{j+N+1})\|_2  +\sum\nolimits_{q=j+1}^{j+N} 2\|\mathrm{wd}(\mathcal{R}_\mathcal{U}^{q})\|_2\Big)$ holds with 
    $\mathrm{wd}(\mathcal{A}) = \overline{\mathcal{A}} - \underline{\mathcal{A}}$ being the width of the interval $\mathcal{A}$. The interval $\mathcal{R}_\mathcal{U}^{q+1}$ is the over-approximation of the reachable set at time index $t_{q+1}$ from the initial uncertain set $\mathcal{R}_\mathcal{U}^q$ (with $\mathcal{R}_\mathcal{U}^j = \Tilde{x}^j$) and for all  $u^q \in \mathcal{U}$. % Such sets can be computed from~\eqref{eq:nextstate} by replacing $u^q$ with $\mathcal{U}$.
\end{theorem}
% \begin{theorem}[\textsc{Suboptimality bound}]\label{thm:suboptimality-gap}
%     Assume that $L_c$ with the $2$-norm is the Lipschitz constant of the cost $c$ on $\mathcal{X} \times \mathcal{U} \times \mathcal{X}$. Let $C_j^{\star}$ and $\hat{C}_j$ be the optimal costs of the $N$-step control problem~\eqref{eq:n-step-optimal-control} when the dynamics are known, e.g. $x^{q+1} = x(\cdot, x^q, u^q)$ is known, and the dynamics are unknown, e.g., $x^{q+1} = h^\theta(x^q,u^q) \in \mathcal{R}^{q+1}$. Then, the following inequality holds
%     \begin{align}
%         |  C_j^* - C_j| \leq L_c \Big( \|\mathrm{wd}(\mathcal{R}_\mathcal{U}^{j+N+1})\|_2  +\sum\nolimits_{q=j+1}^{j+N} 2\|\mathrm{wd}(\mathcal{R}_\mathcal{U}^{q})\|_2\Big),\nonumber
%     \end{align}
%     where $\mathrm{wd}(\mathcal{A}) = \overline{\mathcal{A}} - \underline{\mathcal{A}}$ is the width of the interval $\mathcal{A}$. The interval $\mathcal{R}_\mathcal{U}^{q+1}$ is the over-approximation of the reachable set at time index $t_{q+1}$ from the initial uncertain set $\mathcal{R}_\mathcal{U}^q$ (with $\mathcal{R}_\mathcal{U}^j = \Tilde{x}^j$) and all possible values $u^q \in \mathcal{U}$. Such sets can be computed from~\eqref{eq:nextstate} by replacing $u^q$ with $\mathcal{U}$.
% \end{theorem}
% Note that, Theorem~\ref{thm:suboptimality-gap} also specifies that if the same optimization algorithm is used to compute the solutions of the optimal control problem in both known and unknown case, then the solutions returned by the algorithm will satisfy the suboptimality bound above. 

Theorem~\ref{thm:suboptimality-gap} provides that the suboptimality bound is proportional to the width of the over-approximation of the reachable set. Thus, our algorithm achieves near-optimal control with more data along the trajectory and more side information, as the over-approximations become tighter.

%% file: tex/06_numerical_experiments.tex
\section{Numerical Experiments}
In this section, 
%we illustrate the effectiveness of our control algorithm through a series of examples in MuJoCo~\citep{todorov2012mujoco} and a high-physics F-16 aircraft simulator~\citep{heidlauf2018verification}. 
we empirically demonstrate that the algorithm, using data from only the current trial and the least amount of side information necessary to learn, can achieve performance comparable to the highly-tuned implementations of \dpg{}~\citep{hoffman2020acme} and \sac{}~\citep{pytorch_sac} trained over \emph{ten million} of interactions with the environments. We emphasize that the comparison is unfair to our algorithm since, at each evaluating episode, it learns from only the \emph{thousand} data obtained during the episode. Further, we show in an F-16 aircraft simulator, a $13$-states and $4$-control inputs nonlinear dynamics with polynomial control, that (a) The algorithm outperforms system identification approaches such as \texttt{SINDYc} \citep{kaiser2018sparse} ; (b) The algorithm can meet real-time requirements. We provide further details on the numerical experiments in the extended paper~\citep{djeumou2021learning}. A video of the simulations is at \url{https://tinyurl.com/hdem8x76}, and the code at \url{https://github.com/wuwushrek/datacontrolreach.git}.

\paragraph{Experiments in MuJoCo.}
The equations of motion for multi-joint dynamical systems in the MuJoCo environment are as follows: $M(q) \ddot{q} + b(\dot{q}, q) = h(u) + J_c^{\mathrm{T}}(q) F_c(\dot{q}, q, u)$, where $q$ is the system's state, $M(q)$ is the inertial matrix, $b(\dot{q}, q)$ contains coriolis, centrifugal, gravitational and passive forces,  $J_c^{\mathrm{T}}(q)$ is the contact Jacobian matrix, and $F_c(\dot{q}, q, u)$ is the contact force. 
%We seek to achieve high performance in control tasks such as driving to a target a two-link planar reacher (Reacher) and a $3$-link planar swimmer (Swimmer) based on highly-accurate Reynolds fluid drag model. Besides, we seek to maximize the speed of a running planar biped (Cheetah). 

For each environment, the cost function is provided by MuJoCo, and we perform numeric differentiation in order to find its gradient. The Lipschitz bounds are under-estimated using only $1000$ data points obtained prior to the on-the-fly control. The Reacher environment does not consider any side information other than the Lipschitz bounds, while Swimmer and Cheetah consider that $M(q)$ is known (Side information~\ref{side:partknowledge}) in order to start learning. Indeed, without such side information, our algorithm fails to learn to control due to the large over-approximations of reachable sets. $M(q)$ is typically obtained for a robot through Euler-Lagrange formulation that uses the kinetic and potential energy. Further, we reduce the over-approximation of the contact force $F_c$ by considering the Coulomb law of friction. That is, via Side information~\ref{sideinfo:algebraic}, we impose the constraints $F^1_c \geq 0$ and $F^1_c \geq \sqrt{ (F^2_c)^2 \mu_1 +  (F^3_c)^2 \mu_2}$ at each contact point, where $F^1_c$ is the normal force value, $F^2_c$ and $F^3_c$ are the tangential forces, and $\mu_1, \mu_2$ are the friction coefficients.

Figure $2$ demonstrates that it is possible to learn to control with only data from a single episode by leveraging side information. In Cheetah, more side information can improve our algorithm's performance. We reduced the time-step value of Reacher to accommodate our algorithm and observed that \dpg{} was unable to learn the task solely due to such a change, while \sac{} was not affected.
\begin{figure*}[t]
	\centering
	\label{fig:exp-mujoco}
	\vspace*{-4mm}
	\input{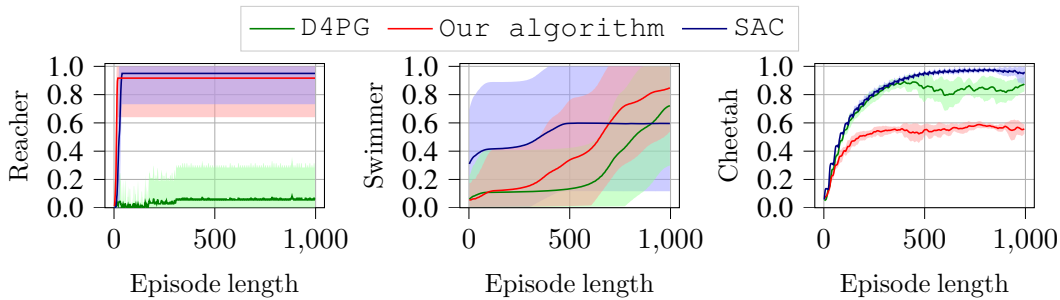}
	\caption{From left to right, we plot the (average) \emph{immediate reward} over $100$ episodes. The experiments show that our algorithm can yield performance comparable to \dpg{} and \sac{}.}
	\vspace*{-8mm}
\end{figure*}

\paragraph{Data-driven control of an F-16 aircraft.}
We consider a scenario involving an F-16 aircraft~\citep{heidlauf2018verification} diving towards the ground at a low altitude and a high downward pitch angle. We show how our algorithm can prevent a ground collision using only the measurements obtained during the dive and elementary laws of physics as side information. We compare our algorithm with the linear-quadratic regulator (LQR) of the simulator, a pre-trained neural network for the task, and \texttt{SINDYc} achieving sparse system identification from a library of functions.

\begin{wrapfigure}{r}{0.55\textwidth}\vspace{-0.4cm}\hspace{-0.20cm}
% 	\hspace*{-1.5em}
	\centering
	\label{f16datacontrol}
	\input{figs/simu_plot_bis}
	\vspace*{-7mm}
	\caption{Our algorithm enables the F-16 to avoid the ground collision while the embedded \texttt{LQR} controller and \texttt{SINDYc} fail to avoid the crash. Further, it can be applied in real time since the compute time is less than the \emph{control time step} enforced by the simulator.}
	\vspace*{-6mm}
\end{wrapfigure}
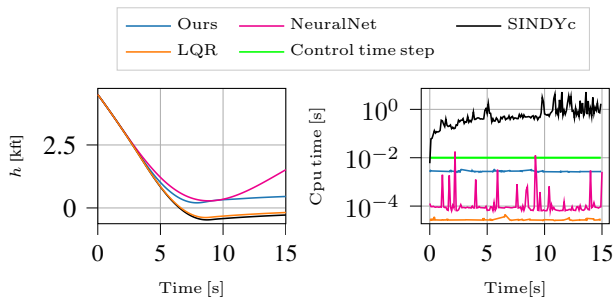
% The F-16's flight control system~\citep{heidlauf2018verification} is described by a hierarchical feedback control loop consisting of an autopilot and a low-level controller. Inside the simulator, the underlying dynamics of the F-16 capture the ($6$-DOF) movement of an aircraft through the standard aerodynamic equations~\citep{stevens2015aircraft}. The dynamics describe the evolution of the system's states, namely velocity $v_t$, angle of attack $\alpha$, sideslip $\beta$, altitude $h$, attitude angles: roll $\phi$, pitch $\theta$, yaw $\psi$, and their corresponding rates $p$, $q$, $r$, engine $power$ and two more states for translation along north and east. The plant model uses several \emph{linearly interpolated lookup tables} to incorporate wind tunnel data describing the engine model, the various coefficients including force and moment coefficients, and the moments due to the control surfaces. We refer the reader to the extended paper~\citep{djeumou2021learning} for futher information.

Our algorithm considers the structural knowledge of rigid-body dynamics while assuming that the aerodynamics forces and moments are completely unknown. In other words, the effect of the control inputs on the aircraft is unknown. For example, from the first principles, the lateral velocity's derivative is given by $rv - q w - g \sin \theta + F_u / m$, where the structure is generic but the aerodynamic force $F_u$ (specific to the aircraft) is unknown. We use the library \texttt{PySINDY}~\citep{desilva2020pysindy} for the comparison with system identification. We considered monomials (up to degree 6), sines and cosines of the state, and the products of these functions with the control inputs as the library functions. We provide the noisy measurements of the state and its derivatives to both \texttt{SINDYc} and our algorithm. Our algorithm uses Lipschitz bounds estimated using $1000$ data points.
Finally, the neural network baseline was trained via policy optimization. Figure $3$ empirically demonstrates the effectiveness of the proposed approach.
%With known look-up tables, the resulting dynamics have \emph{polynomial dependency} in the input.
% \begin{figure}[!hbt]
% % 	\hspace*{-1.5em}
% 	\centering
% 	\label{f16datacontrol}
% 	\input{figs/simu_plot}
% 	\vspace*{-1.em}
% 	\caption{Our algorithm enables the F-16 to avoid the ground collision while the embedded \texttt{LQR} controller and \texttt{SINDYc} fail to avoid the crash. Further, it can be applied in real time since the compute time is less than the \emph{control time step} enforced by the simulator}
% 	\vspace*{-1em}
% \end{figure}

%% file: figs/simu_plot_bis.tex
% This file was created by tikzplotlib v0.9.6.
\pgfplotsset{
	every tick label/.append style={font=\footnotesize},
	width=2.5cm,
	height=1.8cm,
	scale only axis,
	yticklabel style={/pgf/number format/fixed, font=\small},
	yticklabel style={/pgf/number format/fixed, font=\small},
}

\begin{tikzpicture}

\definecolor{color0}{rgb}{0.12156862745098,0.466666666666667,0.705882352941177}
\definecolor{color1}{rgb}{1,0.498039215686275,0.0549019607843137}

\begin{groupplot}[group style={group size=2 by 1, horizontal sep=1.8cm}]
\nextgroupplot[
legend cell align={left},
legend columns=3,
legend style={at={(0.1,1.6)}, fill opacity=1.0, anchor=north west,draw opacity=1, text opacity=1, draw=white!80!black, /tikz/every even column/.append style={column sep=0.2cm}, font=\tiny},
tick align=outside,
tick pos=left,
x grid style={white!69.0196078431373!black},
xmajorgrids,
xmin=0, xmax=15.0099999999997,
xtick style={color=black},
xtick={0, 5, 10, 15},
xticklabels={0, 5, 10, 15},
xlabel={\tiny $\mathrm{Time \ [s]}$},
y grid style={white!69.0196078431373!black},
ylabel={\tiny $h$ [kft]},
ymajorgrids,
ymin=-500.581963509203, ymax=3795.26580778615,
ytick style={color=black},
ytick={-510,0,2000,4000},
yticklabels={−0.5,0,2.5,5}
]
\addplot [semithick, color0]
table {%
0 3600
0.319999933242798 3426.2333984375
0.670000076293945 3233.45751953125
1.02999997138977 3032.38330078125
1.39999997615814 2822.919921875
1.78999996185303 2599.2978515625
2.15000009536743 2390.22314453125
2.51999998092651 2172.38208007812
2.91000008583069 1939.69995117188
3.53999996185303 1563.02099609375
3.75 1441.01281738281
3.9300000667572 1339.3076171875
4.09000015258789 1251.62634277344
4.23999977111816 1172.02490234375
4.3899998664856 1095.08166503906
4.53000020980835 1025.74304199219
4.67000007629395 958.849609375
4.80999994277954 894.406127929688
4.94000005722046 836.75732421875
5.07000017166138 781.211730957031
5.19999980926514 727.787292480469
5.32999992370605 676.491943359375
5.44999980926514 631.063903808594
5.57000017166138 587.514282226562
5.69000005722046 545.871398925781
5.80000019073486 509.409637451172
5.90999984741211 474.619812011719
6.01000022888184 444.473999023438
6.1100001335144 415.762603759766
6.21000003814697 388.515167236328
6.30000019073486 365.262359619141
6.3899998664856 343.227722167969
6.48000001907349 322.426910400391
6.57000017166138 302.873138427734
6.65000009536743 286.549102783203
6.73000001907349 271.226135253906
6.80999994277954 256.910491943359
6.88000011444092 245.217056274414
6.94999980926514 234.295303344727
7.01999998092651 224.154067993164
7.09000015258789 214.78662109375
7.15000009536743 207.377227783203
7.21000003814697 200.544799804688
7.26999998092651 194.28385925293
7.32999992370605 188.58056640625
7.38000011444092 184.261871337891
7.42999982833862 180.339508056641
7.48000001907349 176.81852722168
7.53000020980835 173.68571472168
7.57000017166138 171.454711914062
7.6100001335144 169.475112915039
7.65000009536743 167.745941162109
7.69000005722046 166.2666015625
7.71999979019165 165.320602416992
7.75 164.514526367188
7.78000020980835 163.847885131836
7.80999994277954 163.320526123047
7.82999992370605 163.046051025391
7.84999990463257 162.833190917969
7.86999988555908 162.681793212891
7.88000011444092 162.629058837891
7.8899998664856 162.591751098633
7.90000009536743 162.56965637207
7.90999984741211 162.562957763672
7.92000007629395 162.571441650391
7.92999982833862 162.595275878906
7.94000005722046 162.634857177734
7.94999980926514 162.690673828125
7.96999979019165 162.850448608398
7.98999977111816 163.072723388672
8.01000022888184 163.356719970703
8.02999973297119 163.700576782227
8.0600004196167 164.325973510742
8.09000015258789 165.078170776367
8.11999988555908 165.951751708984
8.15999984741211 167.296173095703
8.19999980926514 168.833648681641
8.23999977111816 170.550628662109
8.28999996185303 172.927856445312
8.34000015258789 175.535934448242
8.39999961853027 178.931823730469
8.47000026702881 183.202209472656
8.55000019073486 188.397262573242
8.65999984741211 195.895156860352
8.96000003814697 216.567504882812
9.0600004196167 223.017471313477
9.14000034332275 227.887268066406
9.22000026702881 232.458404541016
9.28999996185303 236.194427490234
9.35999965667725 239.676208496094
9.43000030517578 242.903671264648
9.5 245.8828125
9.56999969482422 248.625076293945
9.64000034332275 251.145858764648
9.71000003814697 253.463882446289
9.77999973297119 255.59992980957
9.85000038146973 257.576141357422
9.92000007629395 259.414733886719
10 261.375915527344
10.0799999237061 263.218109130859
10.1800003051758 265.398406982422
10.3100004196167 268.110015869141
10.6300001144409 274.7294921875
10.7600002288818 277.547180175781
10.8900003433228 280.479797363281
11.0200004577637 283.528839111328
11.1700000762939 287.174926757812
11.3500003814697 291.681884765625
11.9200000762939 306.055999755859
12.0799999237061 309.920623779297
12.2299995422363 313.427856445312
12.3800001144409 316.813751220703
12.5200004577637 319.863098144531
12.6599998474121 322.809112548828
12.8100004196167 325.858825683594
12.960000038147 328.809173583984
13.1199998855591 331.861053466797
13.3000001907349 335.195465087891
13.4899997711182 338.622253417969
13.710000038147 342.494445800781
13.9399995803833 346.452941894531
14.1800003051758 350.498199462891
14.4200000762939 354.459228515625
14.6499996185303 358.172790527344
14.8699998855591 361.643524169922
15.0100002288818 363.808166503906
};
\addlegendentry{$\mathrm{Ours}$}
\addplot [semithick, magenta]
table {%
0 3600
0.299999952316284 3436.97607421875
0.600000023841858 3271.20190429688
0.910000085830688 3097.13891601562
1.26999998092651 2892.07104492188
2.35999989509583 2268.93481445312
2.65000009536743 2107.06372070312
2.91000008583069 1964.70886230469
3.15000009536743 1836.05419921875
3.36999988555908 1720.77551269531
3.57999992370605 1613.39245605469
3.76999998092651 1518.68493652344
3.96000003814697 1426.45239257812
4.1399998664856 1341.47448730469
4.32000017166138 1258.921875
4.48999977111816 1183.26623535156
4.65000009536743 1114.1962890625
4.80000019073486 1051.44702148438
4.94999980926514 990.73193359375
5.09999990463257 932.11328125
5.23999977111816 879.365905761719
5.38000011444092 828.579833984375
5.51000022888184 783.212036132812
5.6399998664856 739.587219238281
5.76000022888184 700.9052734375
5.88000011444092 663.744934082031
6 628.096069335938
6.11999988555908 593.968078613281
6.23000001907349 564.022521972656
6.34000015258789 535.33740234375
6.44999980926514 507.915985107422
6.55999994277954 481.773040771484
6.65999984741211 459.128936767578
6.76000022888184 437.557373046875
6.8600001335144 417.069793701172
6.96000003814697 397.668884277344
7.05000019073486 381.139404296875
7.1399998664856 365.491058349609
7.23000001907349 350.715148925781
7.32000017166138 336.8095703125
7.40000009536743 325.170623779297
7.48000001907349 314.208892822266
7.55999994277954 303.918029785156
7.6399998664856 294.298034667969
7.71999979019165 285.341369628906
7.78999996185303 278.042755126953
7.8600001335144 271.240142822266
7.92999982833862 264.934295654297
8 259.125122070312
8.0600004196167 254.537017822266
8.11999988555908 250.302490234375
8.18000030517578 246.414932250977
8.23999977111816 242.875671386719
8.28999996185303 240.189117431641
8.34000015258789 237.736145019531
8.39000034332275 235.513381958008
8.4399995803833 233.518264770508
8.48999977111816 231.749603271484
8.52999973297119 230.494323730469
8.56999969482422 229.380218505859
8.60999965667725 228.408050537109
8.64999961853027 227.579879760742
8.68000030517578 227.052841186523
8.71000003814697 226.605514526367
8.73999977111816 226.237335205078
8.77000045776367 225.94709777832
8.78999996185303 225.797470092773
8.8100004196167 225.682846069336
8.82999992370605 225.603088378906
8.84000015258789 225.576110839844
8.85000038146973 225.557800292969
8.85999965667725 225.548126220703
8.86999988555908 225.547073364258
8.88000011444092 225.554611206055
8.89000034332275 225.570709228516
8.89999961853027 225.595275878906
8.92000007629395 225.669906616211
8.9399995803833 225.778427124023
8.96000003814697 225.919418334961
8.97999954223633 226.09326171875
9.01000022888184 226.415771484375
9.03999996185303 226.810577392578
9.06999969482422 227.277313232422
9.10000038146973 227.815826416016
9.14000034332275 228.644393920898
9.18000030517578 229.598846435547
9.22000026702881 230.6787109375
9.26000022888184 231.882293701172
9.3100004196167 233.564971923828
9.35999965667725 235.444671630859
9.40999984741211 237.517700195312
9.46000003814697 239.783004760742
9.52000045776367 242.752075195312
9.57999992370605 245.989730834961
9.64000034332275 249.491302490234
9.71000003814697 253.905090332031
9.77999973297119 258.668853759766
9.85000038146973 263.778167724609
9.92000007629395 269.230377197266
10 275.879791259766
10.0799999237061 282.967315673828
10.1599998474121 290.484069824219
10.25 299.44140625
10.3400001525879 308.915618896484
10.4300003051758 318.888000488281
10.5299997329712 330.551605224609
10.6300001144409 342.810729980469
10.7399997711182 356.956512451172
10.8500003814697 371.778839111328
10.9700002670288 388.68359375
11.0900001525879 406.314575195312
11.2200002670288 426.184143066406
11.3500003814697 446.81201171875
11.4899997711182 469.835205078125
11.6300001144409 493.661071777344
11.7799997329712 520.0380859375
11.9399995803833 549.099975585938
12.1000003814697 579.067321777344
12.2700004577637 611.829162597656
12.4499998092651 647.48388671875
12.6400003433228 686.116882324219
12.8500003814697 729.892578125
13.0799999237061 778.948059082031
13.3299999237061 833.371643066406
13.6199998855591 897.660705566406
13.9700002670288 976.437194824219
14.4399995803833 1083.45532226562
15.0100002288818 1213.43078613281
};
\addlegendentry{$\mathrm{Neural Net}$}
\addplot [semithick, black]
table {%
0 3600
0.299999952316284 3436.98022460938
0.610000014305115 3265.71020507812
0.930000066757202 3086.04125976562
1.25 2903.58276367188
1.58000004291534 2712.62329101562
1.91999995708466 2513.01953125
2.26999998092651 2304.62963867188
2.63000011444092 2087.33959960938
3 1861.02917480469
3.38000011444092 1625.56909179688
3.98000001907349 1252.91223144531
4.19999980926514 1120.27038574219
4.40000009536743 1002.65856933594
4.59000015258789 893.810119628906
4.76999998092651 793.463134765625
4.94000005722046 701.359436035156
5.09999990463257 617.217346191406
5.26000022888184 535.704040527344
5.40999984741211 461.806488037109
5.55000019073486 395.141693115234
5.69000005722046 330.794311523438
5.82000017166138 273.191192626953
5.94999980926514 217.721221923828
6.07999992370605 164.444122314453
6.19999980926514 117.260429382324
6.32000017166138 72.0351943969727
6.42999982833862 32.3329238891602
6.53999996185303 -5.66189336776733
6.65000009536743 -41.9223747253418
6.75 -73.3602676391602
6.84999990463257 -103.326393127441
6.94999980926514 -131.804321289062
7.05000019073486 -158.77880859375
7.1399998664856 -181.758804321289
7.23000001907349 -203.500244140625
7.32000017166138 -223.994567871094
7.40000009536743 -241.158447265625
7.48000001907349 -257.325866699219
7.55999994277954 -272.492340087891
7.6399998664856 -286.654022216797
7.71000003814697 -298.218536376953
7.78000020980835 -309.009124755859
7.84999990463257 -319.024169921875
7.92000007629395 -328.262298583984
7.98000001907349 -335.561553955078
8.03999996185303 -342.288757324219
8.10000038146973 -348.443572998047
8.15999984741211 -354.02587890625
8.21000003814697 -358.240509033203
8.26000022888184 -362.057647705078
8.3100004196167 -365.477508544922
8.35999965667725 -368.500366210938
8.39999961853027 -370.633026123047
8.4399995803833 -372.512054443359
8.47999954223633 -374.137664794922
8.51000022888184 -375.190765380859
8.53999996185303 -376.101593017578
8.56999969482422 -376.870391845703
8.60000038146973 -377.497253417969
8.61999988555908 -377.836395263672
8.64000034332275 -378.112548828125
8.65999984741211 -378.325714111328
8.68000030517578 -378.475952148438
8.6899995803833 -378.527496337891
8.69999980926514 -378.563262939453
8.71000003814697 -378.583312988281
8.72000026702881 -378.587646484375
8.72999954223633 -378.576263427734
8.73999977111816 -378.549011230469
8.75 -378.505645751953
8.76000022888184 -378.446105957031
8.77999973297119 -378.278472900391
8.80000019073486 -378.046783447266
8.81999969482422 -377.752716064453
8.84000015258789 -377.398773193359
8.86999988555908 -376.762512207031
8.89999961853027 -376.010559082031
8.9399995803833 -374.850616455078
8.97999954223633 -373.539978027344
9.02999973297119 -371.737457275391
9.10000038146973 -369.008270263672
9.42000007629395 -356.215576171875
9.51000022888184 -352.948181152344
9.60000038146973 -349.863983154297
9.69999980926514 -346.616058349609
9.81999969482422 -342.8955078125
9.97999954223633 -338.1044921875
10.2399997711182 -330.488433837891
10.5 -323.008209228516
10.710000038147 -317.105773925781
10.9099998474121 -311.624847412109
11.1099996566772 -306.284332275391
11.3100004196167 -301.079223632812
11.5100002288818 -296.003631591797
11.710000038147 -291.053161621094
11.9099998474121 -286.224884033203
12.1099996566772 -281.516357421875
12.3100004196167 -276.925140380859
12.5100002288818 -272.448791503906
12.710000038147 -268.084777832031
12.9099998474121 -263.83056640625
13.1099996566772 -259.683715820312
13.3100004196167 -255.641784667969
13.5100002288818 -251.702484130859
13.710000038147 -247.863479614258
13.9099998474121 -244.12255859375
14.1099996566772 -240.477493286133
14.3100004196167 -236.926116943359
14.5100002288818 -233.466339111328
14.710000038147 -230.096069335938
14.9099998474121 -226.813278198242
15.0100002288818 -225.204071044922
};
\addlegendentry{$\mathrm{SINDYc}$}

\addplot [semithick, color1]
table {%
0 3600
0.299999952316284 3436.97900390625
0.610000014305115 3265.69995117188
0.920000076293945 3091.67358398438
1.25 2903.5380859375
1.58000004291534 2712.55639648438
1.91999995708466 2512.92602539062
2.26999998092651 2304.50341796875
2.63000011444092 2087.16577148438
3 1860.802734375
3.39000010490417 1619.08898925781
3.8199999332428 1352.70202636719
4.03999996185303 1219.66674804688
4.25 1095.74633789062
4.44000005722046 986.4931640625
4.61999988555908 885.711608886719
4.78999996185303 793.15283203125
4.96000003814697 703.334106445312
5.11999988555908 621.470031738281
5.26999998092651 547.212036132812
5.42000007629395 475.483795166016
5.55999994277954 410.917602539062
5.69999980926514 348.736602783203
5.82999992370605 293.202789306641
5.96000003814697 239.855087280273
6.07999992370605 192.600845336914
6.19999980926514 147.300399780273
6.32000017166138 103.993659973145
6.42999982833862 66.0792770385742
6.53999996185303 29.898717880249
6.6399998664856 -1.46684885025024
6.73999977111816 -31.360876083374
6.84000015258789 -59.7668571472168
6.94000005722046 -86.6695098876953
7.03000020980835 -109.584907531738
7.11999988555908 -131.261779785156
7.21000003814697 -151.691558837891
7.28999996185303 -168.798095703125
7.36999988555908 -184.908172607422
7.44999980926514 -200.017379760742
7.53000020980835 -214.121810913086
7.59999990463257 -225.636306762695
7.67000007629395 -236.376998901367
7.73999977111816 -246.342239379883
7.80999994277954 -255.530715942383
7.86999988555908 -262.787536621094
7.92999982833862 -269.472442626953
7.98999977111816 -275.585205078125
8.05000019073486 -281.125640869141
8.10000038146973 -285.305541992188
8.14999961853027 -289.088195800781
8.19999980926514 -292.473754882812
8.23999977111816 -294.896484375
8.27999973297119 -297.065490722656
8.31999969482422 -298.98095703125
8.35999965667725 -300.643127441406
8.39000034332275 -301.723754882812
8.42000007629395 -302.662170410156
8.44999980926514 -303.458587646484
8.47999954223633 -304.113098144531
8.5 -304.470733642578
8.52000045776367 -304.765441894531
8.53999996185303 -304.997283935547
8.5600004196167 -305.166290283203
8.56999969482422 -305.227264404297
8.57999992370605 -305.272583007812
8.59000015258789 -305.302185058594
8.60000038146973 -305.316162109375
8.60999965667725 -305.314453125
8.61999988555908 -305.297119140625
8.63000011444092 -305.263977050781
8.64000034332275 -305.214874267578
8.65999984741211 -305.068359375
8.68000030517578 -304.857818603516
8.69999980926514 -304.58447265625
8.72000026702881 -304.250427246094
8.75 -303.6416015625
8.77999973297119 -302.913177490234
8.8100004196167 -302.077331542969
8.85000038146973 -300.818023681641
8.89999961853027 -299.054534912109
8.96000003814697 -296.737030029297
9.05000019073486 -293.031677246094
9.23999977111816 -285.152954101562
9.32999992370605 -281.635772705078
9.42000007629395 -278.308959960938
9.51000022888184 -275.161224365234
9.60999965667725 -271.835723876953
9.72999954223633 -268.013793945312
9.89999961853027 -262.775726318359
10.1599998474121 -254.935806274414
10.4099998474121 -247.537948608398
10.6199998855591 -241.469161987305
10.8199996948242 -235.835296630859
11.0200004577637 -230.346557617188
11.2200002670288 -224.997802734375
11.4200000762939 -219.783081054688
11.6199998855591 -214.697906494141
11.8199996948242 -209.739166259766
12.0200004577637 -204.904312133789
12.2200002670288 -200.190780639648
12.4200000762939 -195.595977783203
12.6199998855591 -191.117248535156
12.8199996948242 -186.751968383789
13.0200004577637 -182.49755859375
13.2200002670288 -178.351516723633
13.4200000762939 -174.311401367188
13.6199998855591 -170.374816894531
13.8199996948242 -166.539428710938
14.0200004577637 -162.802947998047
14.2200002670288 -159.163101196289
14.4200000762939 -155.617691040039
14.6199998855591 -152.164581298828
14.8199996948242 -148.801651000977
15.0100002288818 -145.688491821289
};
\addlegendentry{$\mathrm{LQR}$}

\addplot [semithick, green]
table {%
0 0
};
\addlegendentry{$\mathrm{Control \ time \ step}$}

\nextgroupplot[
legend cell align={left},
legend style={
  fill opacity=0.8,
  draw opacity=1,
  text opacity=1,
  at={(0.8,0.7)},
  anchor=east,
  draw=white!80!black
},
log basis y={10},
tick align=outside,
tick pos=left,
x grid style={white!69.0196078431373!black},
xlabel={\tiny $\mathrm{Time [s]}$},
xmajorgrids,
xmin=-74.45, xmax=1563.45,
xtick style={color=black},
xtick={0, 500, 1000, 1500},
xticklabels={0, 5, 10, 15},
y grid style={white!69.0196078431373!black},
ylabel={\tiny $\mathrm{Cpu \ time \ [s]}$},
ymajorgrids,
ymin=1.86661269109604e-05, ymax=7.35894076769671,
ymode=log,
ytick style={color=black},
% ytick={1e-06,1e-05,0.0001,0.001,0.01,0.1,1},
% yticklabels={
%   \(\displaystyle {10^{-6}}\),
%   \(\displaystyle {10^{-5}}\),
%   \(\displaystyle {10^{-4}}\),
%   \(\displaystyle {10^{-3}}\),
%   \(\displaystyle {10^{-2}}\),
%   \(\displaystyle {10^{-1}}\),
%   \(\displaystyle {10^{0}}\)
% }
]
\addplot [thick, green]
table {%
0 0.01
1489 0.01
};

\addplot [semithick, color0]
table {%
0 0.0025623083114624
1 0.00260426998138428
2 0.00272693634033203
3 0.00280201435089111
4 0.00279631614685059
5 0.00288803577423096
6 0.00294599533081055
7 0.0029165506362915
8 0.00291421413421631
9 0.00284981727600098
170 0.00269200801849365
171 0.00267069339752197
172 0.00266964435577393
173 0.00266320705413818
174 0.00274507999420166
175 0.00275850296020508
176 0.00278055667877197
177 0.00280501842498779
178 0.00291604995727539
179 0.00292477607727051
180 0.00293350219726562
181 0.00296869277954102
182 0.00297086238861084
183 0.00298233032226562
184 0.00289905071258545
185 0.00288920402526855
186 0.00288503170013428
187 0.00285744667053223
188 0.00275118350982666
189 0.0027761697769165
190 0.0027855396270752
191 0.00275115966796875
192 0.00278151035308838
232 0.00267102718353271
233 0.00263223648071289
234 0.00261235237121582
235 0.00258278846740723
236 0.00257420539855957
237 0.00258677005767822
238 0.00253133773803711
239 0.00255639553070068
240 0.00256876945495605
241 0.00258402824401855
242 0.0026500940322876
243 0.00265882015228271
244 0.0026881217956543
245 0.00269670486450195
246 0.00276238918304443
247 0.00281937122344971
248 0.00284202098846436
249 0.00282061100006104
250 0.00283315181732178
251 0.00281720161437988
252 0.00276768207550049
253 0.00274572372436523
254 0.00270590782165527
255 0.00273449420928955
256 0.00265870094299316
257 0.00262253284454346
258 0.00263280868530273
259 0.00264010429382324
260 0.00264785289764404
261 0.00265581607818603
262 0.00266659259796143
263 0.00268230438232422
264 0.00273022651672363
265 0.00269870758056641
266 0.00280082225799561
267 0.00287814140319824
268 0.00295143127441406
269 0.00309123992919922
270 0.00315852165222168
271 0.00316057205200195
272 0.00317544937133789
273 0.00320694446563721
274 0.00326006412506104
275 0.00325489044189453
367 0.0027263879776001
368 0.00284488201141357
369 0.0027658224105835
370 0.0027773380279541
371 0.00289559364318848
372 0.00291576385498047
386 0.00265090465545654
387 0.00263509750366211
388 0.00263400077819824
389 0.0026357889175415
390 0.00263657569885254
391 0.0026728630065918
392 0.002699875831604
393 0.00268208980560303
394 0.00267188549041748
395 0.00265107154846191
396 0.00268347263336182
397 0.00268869400024414
426 0.0026820182800293
427 0.00264692306518555
428 0.00260481834411621
429 0.00255753993988037
430 0.00258636474609375
431 0.00260591506958008
432 0.00261299610137939
433 0.00263469219207764
434 0.00265464782714844
435 0.00267477035522461
436 0.00268664360046387
437 0.00269424915313721
438 0.00269327163696289
439 0.00269887447357178
440 0.00265154838562012
441 0.00265688896179199
442 0.00266673564910889
443 0.00264720916748047
444 0.00260608196258545
445 0.0025670051574707
446 0.00255260467529297
447 0.00257508754730225
448 0.00258281230926514
449 0.00262513160705566
450 0.00265390872955322
451 0.00266985893249512
452 0.00267000198364258
453 0.00267348289489746
454 0.00268325805664062
455 0.00268058776855469
456 0.00266695022583008
457 0.00268950462341309
458 0.00271370410919189
548 0.00270495414733887
549 0.00273110866546631
550 0.00274326801300049
551 0.00267016887664795
552 0.00269725322723389
553 0.00276517868041992
554 0.00279507637023926
555 0.00282914638519287
556 0.00284996032714844
557 0.00287680625915527
558 0.00315163135528564
559 0.00318853855133057
560 0.0032179594039917
561 0.0032503604888916
562 0.00321907997131348
563 0.00313708782196045
564 0.00315024852752686
565 0.00312643051147461
566 0.00312623977661133
567 0.00311355590820313
568 0.00282180309295654
609 0.00281016826629639
610 0.00276558399200439
611 0.00274851322174072
612 0.00271813869476318
613 0.00276927947998047
614 0.002762770652771
615 0.00279614925384521
616 0.00280025005340576
617 0.00279791355133057
618 0.00277090072631836
619 0.00276260375976563
620 0.00277743339538574
621 0.00279970169067383
622 0.00280997753143311
623 0.00274558067321777
624 0.00279273986816406
625 0.00275514125823975
626 0.00274062156677246
627 0.00271215438842773
628 0.00271153450012207
629 0.00271334648132324
630 0.00272600650787354
631 0.00275859832763672
632 0.0027632474899292
633 0.00276052951812744
634 0.00271704196929932
635 0.002732253074646
636 0.00274853706359863
637 0.00276882648468018
638 0.00275554656982422
639 0.00278422832489014
640 0.00276939868927002
641 0.00271224975585938
642 0.00268275737762451
643 0.00265758037567139
644 0.00265336036682129
645 0.00265321731567383
646 0.00264036655426025
647 0.00264575481414795
648 0.00265846252441406
649 0.00264713764190674
650 0.00265717506408691
651 0.00276603698730469
652 0.0027921199798584
653 0.00280933380126953
654 0.00282161235809326
655 0.00281982421875
656 0.00280787944793701
657 0.00277876853942871
658 0.00275559425354004
695 0.00292937755584717
696 0.00294873714447021
697 0.00294837951660156
698 0.00292057991027832
699 0.00279643535614014
700 0.00278944969177246
785 0.0028048038482666
786 0.00284643173217773
787 0.0028602123260498
788 0.00284867286682129
789 0.00281531810760498
790 0.00280249118804932
791 0.00282692909240723
792 0.00283832550048828
793 0.0028242826461792
794 0.00287532806396484
795 0.00287067890167236
796 0.00280961990356445
797 0.00279369354248047
798 0.00283904075622559
799 0.00287435054779053
800 0.00286190509796143
801 0.00282478332519531
802 0.00284898281097412
803 0.00285587310791016
804 0.00281040668487549
805 0.00288527011871338
882 0.00313348770141602
883 0.00309543609619141
884 0.00297615528106689
885 0.00297777652740479
886 0.00310254096984863
887 0.00306830406188965
888 0.00298976898193359
889 0.00299344062805176
934 0.00263473987579346
935 0.00265064239501953
936 0.00266017913818359
937 0.00267550945281982
938 0.00268058776855469
939 0.00268547534942627
940 0.00267744064331055
941 0.00267758369445801
942 0.0026437520980835
943 0.00258479118347168
944 0.00260992050170898
945 0.00262050628662109
946 0.00262377262115479
947 0.00260910987854004
948 0.00259132385253906
949 0.0026301383972168
950 0.00263664722442627
951 0.00263729095458984
952 0.00266678333282471
953 0.0026965856552124
954 0.00269603729248047
978 0.00279755592346191
979 0.00283012390136719
980 0.00279583930969238
981 0.00278871059417725
982 0.00278263092041016
983 0.00271542072296143
984 0.00272872447967529
985 0.00272555351257324
986 0.00272531509399414
987 0.00273785591125488
988 0.00272459983825684
989 0.00267527103424072
990 0.00267066955566406
991 0.0026759147644043
992 0.0027184009552002
993 0.00271000862121582
994 0.00264642238616943
995 0.00264675617218018
1134 0.00260293483734131
1135 0.00266242027282715
1136 0.00267338752746582
1137 0.00268325805664062
1138 0.00269813537597656
1139 0.00265052318572998
1140 0.00267493724822998
1141 0.00266506671905518
1142 0.00267467498779297
1143 0.00266082286834717
1144 0.0026503324508667
1145 0.00262236595153809
1146 0.00261552333831787
1147 0.00263314247131348
1148 0.00267179012298584
1149 0.00268774032592773
1150 0.00265414714813232
1151 0.00265724658966064
1152 0.00266151428222656
1153 0.00265686511993408
1154 0.00266611576080322
1201 0.00271556377410889
1202 0.00268394947052002
1203 0.00261664390563965
1204 0.00260465145111084
1205 0.00258998870849609
1206 0.00255682468414307
1207 0.00255770683288574
1208 0.00257556438446045
1209 0.00257000923156738
1210 0.00258369445800781
1211 0.0025676965713501
1212 0.00256698131561279
1213 0.00262084007263184
1214 0.00261268615722656
1215 0.00264253616333008
1305 0.00260353088378906
1306 0.00264945030212402
1307 0.002652907371521
1308 0.00268933773040771
1309 0.00271022319793701
1310 0.00267884731292725
1311 0.00266983509063721
1312 0.00266647338867188
1313 0.00266194343566895
1314 0.00266635417938232
1485 0.00267515182495117
1486 0.00264585018157959
1487 0.00267789363861084
1488 0.00267083644866943
1489 0.00262706279754639
};
% \addlegendentry{Our algorithm}
\addplot [semithick, color1]
table {%
0 2.55584716796875e-05
1 2.56061553955078e-05
2 2.57730484008789e-05
3 2.58684158325195e-05
4 2.62022018432617e-05
5 2.6392936706543e-05
6 2.66313552856445e-05
7 2.67744064331055e-05
8 2.68459320068359e-05
9 2.67267227172852e-05
10 2.84671783447266e-05
11 2.84433364868164e-05
12 2.84671783447266e-05
13 2.86340713500977e-05
14 2.84910202026367e-05
15 2.84433364868164e-05
16 2.83718109130859e-05
17 2.83956527709961e-05
18 2.83956527709961e-05
19 2.83479690551758e-05
20 2.65359878540039e-05
21 2.65598297119141e-05
22 2.64883041381836e-05
23 2.64167785644531e-05
24 2.65359878540039e-05
25 2.66313552856445e-05
26 2.66551971435547e-05
27 2.65598297119141e-05
28 2.65359878540039e-05
29 2.65359878540039e-05
30 2.65598297119141e-05
31 2.85863876342773e-05
32 2.86579132080078e-05
33 3.01837921142578e-05
34 3.03506851196289e-05
35 3.34024429321289e-05
36 3.33547592163086e-05
37 3.33786010742188e-05
38 3.34739685058594e-05
39 3.34501266479492e-05
40 3.36408615112305e-05
41 3.16143035888672e-05
42 3.16619873046875e-05
43 3.00884246826172e-05
44 2.97307968139648e-05
45 2.65598297119141e-05
46 2.66551971435547e-05
47 2.68220901489258e-05
48 2.67982482910156e-05
49 2.68936157226562e-05
50 2.67505645751953e-05
51 2.67982482910156e-05
99 2.63214111328125e-05
100 2.61783599853516e-05
101 2.62022018432617e-05
102 2.64883041381836e-05
103 2.66551971435547e-05
104 2.6702880859375e-05
105 2.67505645751953e-05
106 2.68220901489258e-05
107 2.68459320068359e-05
108 2.68936157226562e-05
109 2.70366668701172e-05
110 2.70843505859375e-05
111 2.70605087280273e-05
112 2.67267227172852e-05
113 2.65121459960938e-05
114 2.65359878540039e-05
115 2.66790390014648e-05
116 2.66075134277344e-05
117 2.65598297119141e-05
118 2.65598297119141e-05
119 2.64167785644531e-05
120 2.65598297119141e-05
121 2.6702880859375e-05
122 2.67505645751953e-05
123 2.67505645751953e-05
124 2.66551971435547e-05
125 2.65836715698242e-05
126 2.66551971435547e-05
127 2.67505645751953e-05
128 2.67982482910156e-05
129 2.68220901489258e-05
130 2.66551971435547e-05
131 2.66551971435547e-05
132 2.66790390014648e-05
133 2.66790390014648e-05
134 2.6702880859375e-05
135 2.66551971435547e-05
136 2.65598297119141e-05
137 2.66551971435547e-05
138 2.67505645751953e-05
139 2.67982482910156e-05
140 2.68697738647461e-05
141 2.67982482910156e-05
142 2.67982482910156e-05
143 2.69412994384766e-05
144 2.69651412963867e-05
145 2.69412994384766e-05
146 2.69889831542969e-05
147 2.68220901489258e-05
148 2.68220901489258e-05
149 2.83718109130859e-05
150 2.83718109130859e-05
151 2.83718109130859e-05
152 2.83479690551758e-05
153 2.81810760498047e-05
154 2.81572341918945e-05
155 2.81572341918945e-05
156 2.82049179077148e-05
157 2.83479690551758e-05
158 2.82049179077148e-05
159 2.65836715698242e-05
160 2.65598297119141e-05
161 2.65836715698242e-05
162 2.65836715698242e-05
163 2.66075134277344e-05
164 2.65598297119141e-05
165 2.65359878540039e-05
166 2.64883041381836e-05
309 2.67982482910156e-05
310 2.67744064331055e-05
311 2.68220901489258e-05
312 2.70605087280273e-05
313 2.7155876159668e-05
314 2.72035598754883e-05
315 2.7155876159668e-05
316 2.70366668701172e-05
317 2.70843505859375e-05
461 2.63690948486328e-05
462 2.82526016235352e-05
463 2.82526016235352e-05
464 3.00884246826172e-05
465 3.01361083984375e-05
466 3.01599502563477e-05
467 3.01599502563477e-05
468 3.01837921142578e-05
469 3.0207633972168e-05
470 3.02553176879883e-05
471 3.02791595458984e-05
472 2.84910202026367e-05
473 2.8538703918457e-05
474 2.69651412963867e-05
475 2.7155876159668e-05
476 2.71081924438477e-05
477 2.70605087280273e-05
478 2.7012825012207e-05
479 2.69651412963867e-05
480 2.69174575805664e-05
481 2.69174575805664e-05
522 2.64883041381836e-05
523 2.65836715698242e-05
524 2.84194946289063e-05
525 2.84671783447266e-05
526 2.84433364868164e-05
527 2.84194946289063e-05
528 2.8228759765625e-05
529 2.8228759765625e-05
530 2.83002853393555e-05
531 2.82526016235352e-05
532 2.83002853393555e-05
533 2.81333923339844e-05
534 2.62737274169922e-05
535 2.61545181274414e-05
536 2.61068344116211e-05
537 2.61068344116211e-05
538 2.61306762695312e-05
539 2.60591506958008e-05
540 2.58922576904297e-05
541 2.60114669799805e-05
542 2.59876251220703e-05
543 2.59876251220703e-05
544 2.59876251220703e-05
545 2.61068344116211e-05
546 2.62260437011719e-05
547 2.6249885559082e-05
548 2.7012825012207e-05
645 3.5548210144043e-05
646 3.5548210144043e-05
647 3.56197357177734e-05
648 3.64780426025391e-05
649 3.63826751708984e-05
650 3.63588333129883e-05
651 3.64065170288086e-05
652 3.65257263183594e-05
653 3.66926193237305e-05
654 3.9982795715332e-05
655 4.35352325439453e-05
656 4.35590744018555e-05
657 4.35113906860352e-05
658 4.27722930908203e-05
659 4.26769256591797e-05
699 2.62975692749023e-05
700 2.62737274169922e-05
701 2.63690948486328e-05
702 2.65359878540039e-05
703 2.64883041381836e-05
704 2.64883041381836e-05
705 2.64406204223633e-05
706 2.63690948486328e-05
707 2.63452529907227e-05
708 2.61545181274414e-05
709 2.60114669799805e-05
710 2.58684158325195e-05
711 2.58207321166992e-05
712 2.56538391113281e-05
713 2.57492065429688e-05
714 2.59637832641602e-05
769 2.77996063232422e-05
770 2.77996063232422e-05
771 2.78234481811523e-05
772 2.77996063232422e-05
773 2.593994140625e-05
774 2.59637832641602e-05
802 2.82049179077148e-05
803 2.81810760498047e-05
804 2.83002853393555e-05
805 2.82764434814453e-05
806 2.69889831542969e-05
807 2.70605087280273e-05
808 2.68459320068359e-05
809 2.68459320068359e-05
810 2.68220901489258e-05
811 2.66551971435547e-05
812 2.66551971435547e-05
813 2.63690948486328e-05
814 2.62022018432617e-05
815 2.61306762695312e-05
816 2.59160995483398e-05
817 2.57492065429688e-05
869 2.85148620605469e-05
870 2.84433364868164e-05
871 2.85863876342773e-05
872 2.67505645751953e-05
873 2.6702880859375e-05
874 2.66313552856445e-05
982 2.62022018432617e-05
983 2.61306762695312e-05
984 2.61545181274414e-05
985 2.54392623901367e-05
986 2.55346298217773e-05
987 2.56061553955078e-05
988 2.57015228271484e-05
989 2.57968902587891e-05
990 2.56538391113281e-05
991 2.57015228271484e-05
992 2.58922576904297e-05
993 2.60829925537109e-05
994 2.61783599853516e-05
995 2.65359878540039e-05
996 2.68220901489258e-05
997 2.68936157226562e-05
998 2.71320343017578e-05
999 2.72274017333984e-05
1000 2.73942947387695e-05
1001 2.74896621704102e-05
1002 2.72989273071289e-05
1068 2.67744064331055e-05
1069 2.67505645751953e-05
1070 2.67982482910156e-05
1071 2.68459320068359e-05
1072 2.67744064331055e-05
1073 2.66790390014648e-05
1074 2.66075134277344e-05
1075 2.66313552856445e-05
1076 2.64883041381836e-05
1077 2.64406204223633e-05
1078 2.63214111328125e-05
1079 2.62737274169922e-05
1080 2.60114669799805e-05
1081 2.56776809692383e-05
1082 2.55584716796875e-05
1083 2.54392623901367e-05
1084 2.52962112426758e-05
1085 2.52246856689453e-05
1086 2.52246856689453e-05
1087 2.52008438110352e-05
1088 2.5177001953125e-05
1089 2.52008438110352e-05
1090 2.53915786743164e-05
1091 2.55107879638672e-05
1092 2.55823135375977e-05
1093 2.57492065429688e-05
1094 2.58445739746094e-05
1095 2.57015228271484e-05
1096 2.57968902587891e-05
1097 2.58207321166992e-05
1153 2.61306762695312e-05
1154 2.63452529907227e-05
1155 2.62975692749023e-05
1156 2.62737274169922e-05
1157 2.63690948486328e-05
1158 2.64167785644531e-05
1159 2.64167785644531e-05
1160 2.63690948486328e-05
1161 2.63214111328125e-05
1162 2.64883041381836e-05
1163 2.65359878540039e-05
1164 2.63690948486328e-05
1165 2.63214111328125e-05
1207 3.03268432617188e-05
1208 3.19242477416992e-05
1209 3.19719314575195e-05
1210 3.20196151733398e-05
1211 3.20911407470703e-05
1212 3.21388244628906e-05
1213 3.21626663208008e-05
1214 3.23295593261719e-05
1215 3.24726104736328e-05
1216 3.03506851196289e-05
1217 2.83479690551758e-05
1218 2.68220901489258e-05
1219 2.67267227172852e-05
1220 2.66313552856445e-05
1221 2.65598297119141e-05
1222 2.6392936706543e-05
1223 2.63690948486328e-05
1224 2.63214111328125e-05
1225 2.62737274169922e-05
1226 2.62975692749023e-05
1227 2.62022018432617e-05
1228 2.60829925537109e-05
1229 2.61783599853516e-05
1230 2.62975692749023e-05
1231 2.63452529907227e-05
1232 2.6392936706543e-05
1233 2.63452529907227e-05
1234 2.63452529907227e-05
1235 2.63214111328125e-05
1236 2.61306762695312e-05
1237 2.60114669799805e-05
1238 2.58684158325195e-05
1239 2.5629997253418e-05
1240 2.55107879638672e-05
1241 2.53438949584961e-05
1242 2.53677368164063e-05
1243 2.53200531005859e-05
1296 2.65121459960938e-05
1297 2.6392936706543e-05
1298 2.62975692749023e-05
1299 2.62737274169922e-05
1300 2.62022018432617e-05
1301 2.62260437011719e-05
1302 2.62022018432617e-05
1303 2.61783599853516e-05
1304 2.62260437011719e-05
1305 2.6702880859375e-05
1306 2.63690948486328e-05
1307 2.6392936706543e-05
1308 2.65121459960938e-05
1309 2.65359878540039e-05
1310 2.66313552856445e-05
1351 2.65836715698242e-05
1352 2.67982482910156e-05
1353 2.69889831542969e-05
1354 2.68936157226562e-05
1355 2.67267227172852e-05
1356 2.66551971435547e-05
1357 2.66790390014648e-05
1358 2.67744064331055e-05
1359 2.87055969238281e-05
1360 2.85148620605469e-05
1361 2.82526016235352e-05
1362 2.81333923339844e-05
1363 2.82526016235352e-05
1364 2.83479690551758e-05
1365 2.83718109130859e-05
1366 2.84194946289063e-05
1367 2.83956527709961e-05
1368 2.82049179077148e-05
1369 2.63214111328125e-05
1370 2.64406204223633e-05
1371 3.08752059936523e-05
1372 3.08036804199219e-05
1373 3.06129455566406e-05
1374 3.04937362670898e-05
1375 3.05891036987305e-05
1376 3.06129455566406e-05
1377 3.06129455566406e-05
1378 3.06367874145508e-05
1379 3.05891036987305e-05
1423 2.95877456665039e-05
1424 2.96115875244141e-05
1425 2.95639038085938e-05
1426 2.9754638671875e-05
1427 2.9754638671875e-05
1428 2.72035598754883e-05
1429 2.69651412963867e-05
1430 2.6702880859375e-05
1431 2.66075134277344e-05
1432 2.66313552856445e-05
1433 2.66075134277344e-05
1434 2.65598297119141e-05
1435 2.65121459960938e-05
1436 2.63214111328125e-05
1476 2.70366668701172e-05
1477 2.8228759765625e-05
1478 2.8228759765625e-05
1479 2.82049179077148e-05
1480 2.80857086181641e-05
1481 2.80141830444336e-05
1482 2.80141830444336e-05
1483 2.79426574707031e-05
1484 2.7918815612793e-05
1485 2.79426574707031e-05
1486 2.78949737548828e-05
1487 2.67267227172852e-05
1488 2.6702880859375e-05
1489 2.68220901489258e-05
};

\addplot [semithick, black]
table {%
0 0.00592088745906949
10 0.0600862503051758
20 0.0638511180877686
30 0.0910444185137749
40 0.132396712899208
50 0.104731075465679
60 0.119893081486225
70 0.116265766322613
80 0.120471931993961
90 0.133967623114586
100 0.355528831481934
110 0.335725843906403
120 0.245364919304848
130 0.195368036627769
140 0.396350860595703
150 0.220376014709473
160 0.196357741951942
170 0.202601879835129
180 0.225653186440468
190 0.159676313400269
200 0.181854739785194
210 0.159049734473228
220 0.16515588760376
230 0.229186341166496
240 0.220695033669472
250 0.355020463466644
260 0.307507038116455
270 0.352598190307617
280 0.353427648544312
290 0.414069652557373
300 0.260387629270554
310 0.30751845240593
320 0.296747177839279
330 0.505423307418823
340 0.293178796768188
350 0.350566655397415
360 0.303786516189575
370 0.402692288160324
380 0.781724154949188
390 0.38451412320137
400 0.387077033519745
410 0.42944461107254
420 0.448832809925079
430 0.433414220809937
440 0.398236483335495
450 0.406435936689377
460 0.394680261611938
470 0.405991822481155
480 0.907923579216003
490 1.08585238456726
500 0.685379981994629
510 1.70847427845001
520 0.700983464717865
530 0.679914176464081
540 0.302210122346878
550 0.411785900592804
560 0.325781792402267
570 0.331409186124802
580 0.442776143550873
590 0.383576661348343
600 0.291091680526733
610 0.301702946424484
620 0.466920405626297
630 0.410944670438766
640 0.502810895442963
650 0.516024529933929
660 0.375837355852127
670 0.421245545148849
680 0.415804862976074
690 0.538586556911469
700 0.386828690767288
710 0.476672679185867
720 0.512127697467804
730 0.395776301622391
740 0.442476034164429
750 0.437187910079956
760 0.41150376200676
770 0.416067570447922
780 0.61284464597702
790 0.488933056592941
800 0.613175809383392
810 0.443936139345169
820 0.592047095298767
830 0.483545035123825
840 0.485193252563477
850 0.490949362516403
860 0.503785848617554
870 0.490179270505905
880 0.444564878940582
890 0.513247013092041
900 0.545936286449432
910 0.399695843458176
920 0.532615125179291
930 0.507783889770508
940 0.373887360095978
950 0.463179975748062
960 0.466022968292236
970 0.712039172649384
980 3.1496422290802
990 2.06347870826721
1000 0.647189676761627
1010 0.607597768306732
1020 0.577193319797516
1030 0.616701066493988
1040 0.517397582530975
1050 0.682519435882568
1060 0.76899790763855
1070 0.907562017440796
1080 2.81309819221497
1090 4.26701068878174
1100 0.67614483833313
1110 0.93047422170639
1120 0.547680377960205
1130 2.47098541259766
1140 0.74678099155426
1150 5.24157190322876
1160 0.571841716766357
1170 0.915955483913422
1180 0.622811794281006
1190 0.874329388141632
1200 1.86035561561584
1210 3.57662153244019
1220 1.71514272689819
1230 3.35886406898499
1240 0.852477431297302
1250 0.893048524856567
1260 0.43448281288147
1270 1.123703956604
1280 0.583499908447266
1290 0.79275768995285
1300 0.507150828838348
1310 2.27052140235901
1320 0.668072164058685
1330 1.84204566478729
1340 0.547144412994385
1350 0.751754224300385
1360 5.23997688293457
1370 0.633425533771515
1380 2.22952818870544
1390 0.626921713352203
1400 1.09055173397064
1410 0.522374629974365
1420 3.13471245765686
1430 0.679497539997101
1440 1.21866726875305
1450 0.913140654563904
1460 1.81417489051819
1470 0.774757206439972
1480 0.723509728908539
1490 1.70638394355774
};

\addplot [semithick, magenta]
table {%
0 0.000127792372950353
10 9.53674461925402e-05
20 9.36984943109564e-05
30 9.82284036581405e-05
40 9.22680119401775e-05
50 8.96454002941027e-05
60 9.20295424293727e-05
70 9.1791182057932e-05
80 8.94069307832979e-05
90 9.22680119401775e-05
100 9.01222592801787e-05
110 0.00197458220645785
120 9.29832604015246e-05
130 8.72611999511719e-05
140 9.0599074610509e-05
150 7.03335026628338e-05
160 6.67572094243951e-05
170 0.00186610268428922
180 9.32216426008381e-05
190 6.46114349365234e-05
200 0.000108480460767169
210 8.55922698974609e-05
220 0.0178499203175306
230 6.53266906738281e-05
240 6.43730236333795e-05
250 6.96182178216986e-05
260 8.79764556884766e-05
270 0.000114202499389648
280 8.55922698974609e-05
290 8.72611999511719e-05
300 8.74996112543158e-05
310 8.36848776089028e-05
320 8.72611999511719e-05
330 9.01222592801787e-05
340 8.48770141601562e-05
350 8.63074965309352e-05
360 8.32080841064453e-05
370 9.96590097201988e-05
380 8.41617220430635e-05
390 8.36848776089028e-05
400 0.0012471677036956
410 8.39233762235381e-05
420 8.55922698974609e-05
430 9.53674461925402e-05
440 8.67843191372231e-05
470 8.46385955810547e-05
480 8.34465099615045e-05
490 9.75131697487086e-05
500 6.62803649902344e-05
520 6.81876772432588e-05
530 9.41753023653291e-05
540 9.20295424293727e-05
550 6.55651092529297e-05
560 0.000139236479299143
570 6.69956280034967e-05
580 8.34465099615045e-05
590 0.00320386956445873
600 8.67843191372231e-05
610 8.67843191372231e-05
620 8.58306448208168e-05
630 8.46385955810547e-05
640 8.44001915538684e-05
650 8.34465099615045e-05
660 9.03606487554498e-05
670 9.01222592801787e-05
680 8.8930144556798e-05
690 8.82148815435357e-05
700 8.63074965309352e-05
710 8.51154545671307e-05
720 8.60691216075793e-05
730 8.91685122041963e-05
750 9.01222592801787e-05
760 0.000797510205302387
770 8.34465099615045e-05
780 8.55922698974609e-05
790 9.0599074610509e-05
800 6.65187617414631e-05
810 0.00010156630014535
820 6.67572094243951e-05
830 9.46521540754475e-05
840 0.000454664201242849
850 0.000630140362773091
860 8.77379934536293e-05
870 8.08239055913873e-05
880 6.72340102028102e-05
890 6.74724287819117e-05
900 6.55651092529297e-05
910 8.84532782947645e-05
920 0.0123758306726813
930 0.00143313372973353
940 6.77108764648438e-05
950 6.55651092529297e-05
960 0.000382423488190398
970 6.46114349365234e-05
980 6.43730236333795e-05
990 6.91413588356227e-05
1000 6.48498098598793e-05
1010 6.41345686744899e-05
1020 6.48498098598793e-05
1030 8.98838115972467e-05
1040 0.000651597802061588
1050 6.50882793706842e-05
1060 0.000113725633127615
1070 7.24792625987902e-05
1080 8.74996112543158e-05
1090 7.17639777576551e-05
1100 6.86645216774195e-05
1110 6.67572094243951e-05
1120 6.58035351079889e-05
1130 6.41345686744899e-05
1140 6.55651092529297e-05
1150 6.55651092529297e-05
1160 6.60419536870904e-05
1170 0.000151157364598475
1180 6.41345686744899e-05
1190 6.79492659401149e-05
1200 6.74724287819117e-05
1210 7.17639777576551e-05
1220 7.31945256120525e-05
1230 6.74724287819117e-05
1240 9.48906454141252e-05
1250 6.74724287819117e-05
1260 6.67572094243951e-05
1270 6.58035351079889e-05
1280 9.2506394139491e-05
1290 9.53674461925402e-05
1300 6.81876772432588e-05
1310 0.00010418892634334
1320 7.22408367437311e-05
1330 6.79492659401149e-05
1340 6.67572094243951e-05
1350 7.36713482183404e-05
1360 0.000104427308542654
1370 8.34465099615045e-05
1380 7.15255446266383e-05
1390 7.00950549799018e-05
1400 0.00293421768583357
1410 8.58306448208168e-05
1420 9.41753023653291e-05
1430 0.000137805938720703
1440 8.96454002941027e-05
1450 8.86917332536541e-05
1460 8.70227668201551e-05
1470 8.29697091830894e-05
1480 8.98838115972467e-05
1490 9.32216426008381e-05
1500 0.00260067032650113
};

\end{groupplot}

\end{tikzpicture}

%% file: tex/07_conclusion.tex
\section{Conclusion} \label{sec:conclusion}\vspace{-0.20cm}
This paper develops a learning-based, data-efficient control algorithm for unknown systems using streaming data from an ongoing trial and available side information. 
The experiments demonstrate that it is possible, with data from a single episode and side information, to perform comparably to learning algorithms trained over millions of environment interactions. Further, we empirically show that the algorithm is fast and can be used in a scenario with real-time constraints.

%% file: tex/08_appendix.tex
\clearpage
% \newpage

\begin{center} % supp mat title
        {\LARGE \bf Learning to Reach, Swim, Walk and Fly in One Trial:} \vspace*{0.15cm}\\
        {\Large \bf Control of Unknown Systems with Scarce Data and Side Information}\vspace*{0.15cm} \\
        $-$ \Large \normalfont Supplementary Material $-$\vspace*{0.15cm}
\end{center}

In this supplementary material, we provide the proofs of the lemma and theorems described in the paper, and additional insights on the numerical experiments.

\section*{Proof of Lemma~\ref{lem:overapprox-fg}}
This is a direct result from combining the arithmetic of intervals and the definition of the bounds $\lipf_k$ and $\lipg_{p,l}$ provided by the Lipschitz assumption. Specifically, from the upper bound $\lipf_k$ on the Lipschitz constant of $f_k$, we have that 
\begin{align*}
	   |f_k(x) - f_k(y)| \leq \lipf_k \|x-y\|_w, \: \: \forall x,y \in \mathcal{X}.
\end{align*}
Hence, given $(\Tilde{x}^i, C_{\mathcal{F}^i}, C_{\mathcal{G}^i}) \in \mathscr{E}_j$ and $x \in  \mathcal{X}$, we can write that $f_k(x) \in f_k(\Tilde{x}^i) + [-1,1] \lipf_k \|x-\Tilde{x}^i\|_w$, and therefore $f_k(x) \in C_{\mathcal{F}_k^i} + [-1,1] \lipf_k \|x-\Tilde{x}^i\|_w $ due to $f(\Tilde{x}^i) \in C_{\mathcal{F}^i}$. Now we want to extend the function $\|\cdot\|_w$ to a function $\boldsymbol{\eta}^w(\cdot)$ in the domain of intervals. We have that
\begin{align}
	\boldsymbol{\eta}^w(\mathcal{S}) = \beta_1 \Big( \sum_{i=1}^n w_i \beta_2(\mathcal{S}_i) \Big), \: \forall \mathcal{S} \in  \mathbb{IR}^n,
	\end{align}
	where the functions $\boldsymbol{\beta}_1: \mathbb{IR} \mapsto \mathbb{IR}$ and $\boldsymbol{\beta}_2: \mathbb{IR} \mapsto \mathbb{IR}$ are interval extensions of $\sqrt{\cdot}$ and $(\cdot)^2$, respectively. For any $\mathcal{S} \in \mathbb{IR}$, we have that
	\begin{align}
	\boldsymbol{\beta}_1(\mathcal{S}) &= [\sqrt{\underline{\mathcal{S}}}, \sqrt{\overline{\mathcal{S}}}], \: \: \: \text{if } \underline{\mathcal{S}} \geq 0 , \label{eq:sqrt_ext}\\
	\boldsymbol{\beta}_2(\mathcal{S}) &= \begin{cases} [0 , \max\{\underline{\mathcal{S}}^2,\overline{\mathcal{S}}^2\}],& \text{if } 0 \in \mathcal{S} \\
	[\min \{\underline{\mathcal{S}}^2,\overline{\mathcal{S}}^2\}, \max\{\underline{\mathcal{S}}^2,\overline{\mathcal{S}}^2\}],& \text{otherwise}.\end{cases}\label{eq:sqr_ext}
\end{align}

By monotonicity of $\sqrt{\cdot}$ and $(\cdot)^2$, it is easy to see that $\{\sqrt{a} \vert a \in \mathcal{A}\} \subseteq \boldsymbol{\beta}_1(\mathcal{A})$ and $\{a^2 \vert a \in \mathcal{A}\} \subseteq \boldsymbol{\beta}_2(\mathcal{A})$ for given $\mathcal{A} \in \mathbb{IR}$. Hence, it is immediate that $\boldsymbol{\eta}^w$ is an interval extension of $\|\cdot\|_w$. Thus, for all $x \in \mathcal{A} \subseteq \mathcal{X}$, we have 
\begin{align*}
	   f_k(x) \in f_k(\Tilde{x}^i) + [-1,1] \lipf_k \|x-\Tilde{x}^i\|_w \in f_k(x^i) + [-1,1] \lipf_k \boldsymbol{\eta}^w(\mathcal{A}-\Tilde{x}^i) .
\end{align*}
Therefore, $\mathscr{R}(f_k,\mathcal{A}) = \{f_k(x) | x \in \mathcal{A}\} \subseteq f_k(\Tilde{x}^i) + [-1,1]\lipf_k \boldsymbol{\eta}^w(\mathcal{A}-\Tilde{x}^i)$. The previous belonging relation is valid for every data $(\Tilde{x}^i, C_{\mathcal{F}^i}, C_{\mathcal{G}^i)} \in \mathscr{E}_j$,and, as a result, $\mathscr{R}(f_k,\mathcal{A}) \subseteq \boldsymbol{f}_k(\mathcal{A})$. The same reasoning applied to $\lipg_{p,k}$ enables to show that $\mathscr{R}(g_{p,k},\mathcal{A})  \subseteq \boldsymbol{g}_{p,k}(\mathcal{A})$.

\section*{Proof of Lemma~\ref{lem:contraction}}
    The proof is very similar to the result in~\cite{djeumou2021fly}, proven for the noiseless data setting.
    Given the knowledge that $f(\Tilde{x}^i) \in \mathcal{F}^i$ and $g_{p,k}(\Tilde{x}^i) \in \mathcal{G}_{p,k}^i$, we seek for tighter intervals $C_{\mathcal{F}^i} \subseteq \mathcal{F}^i$ and $C_{\mathcal{G}_{p,k}^i} \subseteq \mathcal{G}_{p,k}^i$ that prune out some values $f_k(\Tilde{x}^i)$ and $g_{p,k}(\Tilde{x}^i)$ from $\mathcal{F}_k^i$ and $\mathcal{G}_{p,k}^i$ that do not satisfy the constraint $\dot{\Tilde{x}}^i = f(\Tilde{x}^i) + \sum_{p=1}^d g_p(\Tilde{x}^i)\textit{u}^i[\alpha^p]$. Note that the proof below can be straightforwardly adapted for general linear constraints in the form $z^\mathrm{T} w =v$ where $v$ and $w$ are known and $z$ is the variable that need to be contracted. We have that $ \dot{\Tilde{x}}^i \in [\Tilde{\dot{x}}_i - \bar{\eta},\Tilde{\dot{x}}^i + \bar{\eta}]$ by the noise bound assumption. Therefore, we have that
    \begin{align*}
        f(\Tilde{x}^i) = \dot{\Tilde{x}}^i - \sum_{p=1}^d g_p(\Tilde{x}^i) \textit{u}^i[\alpha^p] \in ([\Tilde{\dot{x}}_i - \bar{\eta},\Tilde{\dot{x}}^i + \bar{\eta}] - \sum_{p=1}^d \mathcal{G}^i_{p,k} \pU^i[\alpha^p]) \cap \mathcal{F}^i = C_{\mathcal{F}^i}.
    \end{align*}
    Therefore, a similar reasoning using the tighter interval $C_{\mathcal{F}^i}$ and interval arithmetic provides that 
    \begin{align*}
        \sum_{p=1}^d g_p(\Tilde{x}^i) \textit{u}^i[\alpha^p] \in ([\Tilde{\dot{x}}_i - \bar{\eta},\Tilde{\dot{x}}^i + \bar{\eta}] - C_{\mathcal{F}^i}) \cap (\sum_{p=1}^d \mathcal{G}^i_{p,k} \pU^i[\alpha^p]) = \mathcal{S}_0.
    \end{align*}
    
    Note that plugging back $\mathcal{S}_0$ instead of $\sum_{p=1}^d \mathcal{G}^i_{p,k} \pU^i[\alpha^p]$ in the expression of $C_{\mathcal{F}^i}$ will not yield a tighter set. Therefore, $C_{\mathcal{F}^i}$ and $\mathcal{S}_0$ are optimal. Next, we focus on the term $\sum_{p=1}^d g_p(\Tilde{x}^i) \textit{u}^i[\alpha^p] \in \mathcal{S}_0$. For all $k \in \mathbb{N}_{[1,n]}$, we have that
    \begin{align*}
        g_{1,k}(\Tilde{x}^i) \textit{u}^i[\alpha^1] = \sum_{p=1}^d g_{p,k}(\Tilde{x}^i) \textit{u}^i[\alpha^p] - \sum_{p>1}^d g_{p,k}(\Tilde{x}^i) \textit{u}^i[\alpha^p] \in \underbrace{\big( \mathcal{S}_{0,k} - \sum_{l = 2}^d \mathcal{G}^i_{l,k} \pU^i[\alpha^l]\big) \cap (\mathcal{G}^i_{1,k}  \textit{u}^i[\alpha^1]) }_{C_{\mathcal{G}^i_{1,k}} \textit{u}^i[\alpha^1]},
    \end{align*}
    and we can, in a similar manner, deduce that  
    \begin{align*}
        \sum_{p > 1 }^d g_{p,k}(\Tilde{x}^i) \textit{u}^i[\alpha^p] &\in \underbrace{\big( \mathcal{S}_{0,k} - C_{\mathcal{G}^i_{1,k}}  \textit{u}^i[\alpha^1] \big) \cap \big(  \sum_{p > 1 }^d \mathcal{G}^i_{p,k} \textit{u}^i[\alpha^p] \big)}_{\mathcal{S}_{1,k}}.
    \end{align*}
    Using the same argument as for the optimality of $\mathcal{S}_0$ and $C_{\mathcal{F}^i}$, we can say that $\mathcal{S}_{1,k}$ and $C_{\mathcal{G}_{1,k}^i} \textit{u}^i[\alpha^1]$ are optimal. Finally, we apply the previous step in a sequential manner for $p=2,\hdots,d$ to the equality $g_{p,k}(\Tilde{x}^i) \textit{u}^i[\alpha^p] = \sum_{l > p-1 }^d g_{l,k}(\Tilde{x}^i) \textit{u}^i[\alpha^l] - \sum_{l > p }^d g_{l,k}(\Tilde{x}^i) \textit{u}^i[\alpha^l]$ in order to obtain optimal intervals $\mathcal{S}_{p,k}$ and $C_{\mathcal{G}_{p,k}^i} \textit{u}^i[\alpha^p]$.  

\section*{Proof of Theorem~\ref{thm:diffinclusion}}

The result is straightforward from Lemma~\ref{lem:overapprox-fg} and Lemma~\ref{lem:contraction}. First, let $i \in \mathbb{N}_{[1,j]}$. We show that for all $(\Tilde{x}^i, C_{\mathcal{F}^i}, C_{\mathcal{G}^i}) \in \mathscr{E}_j$ given by Algorithm~\ref{algo:construction}, we have $f_k(\Tilde{x}^i) \in C_{\mathcal{F}_k^i}$ and $g_{p,k}(\Tilde{x}^i) \in C_{\mathcal{G}_i}$ for all $p,k\in \Nd \times \Nn$ . Specifically, as a consequence of line~\ref{alg:update-ei} of Algorithm~\ref{algo:refine} and Lemma~\ref{lem:overapprox-fg}, we have that $f_k(\Tilde{x}^i) \in \mathcal{F}_k^i$ and $g_{p,k}(\Tilde{x}^i) \in \mathcal{G}^i_{p,k}$. Hence, by line~\ref{alg:contraction-f-G} of Algorithm~\ref{algo:refine} and Lemma~\ref{lem:contraction}, we immediately have that $f_k(\Tilde{x}^i) \in C_{\mathcal{F}_k^i}$ and $g_{p,k}(\Tilde{x}^i) \in C_{\mathcal{G}_{k,p}^i}$. Thus, $\mathscr{E}_j$ can be used in Lemma~\ref{lem:overapprox-fg} to conclude that $f_k(x) \in \boldsymbol{f}_k(x)$ and $g_{p,k}(x) \in \boldsymbol{g}_{p,k}(x)$ for all $x \in \mathcal{X}$. Therefore, we have $\dot{x} = f(x) + \sum_{p=1}^d g_p(x) \pU[\alpha^p] \in \boldsymbol{f}(x) + \sum_{p=1}^d \boldsymbol{g}_p(x) \pU[\alpha^p]$ through straightforward interval arithmetic.

\section*{Proof of Theorem~\ref{thm:one-step-overapprox}}

    This proof leverages an interval Taylor-based method to over-approximate the reachable set of dynamics described by differential inclusions. 
    
    First, consider the case of known dynamics in the form $\dot{x} = h(x,u)$, where $h : \mathcal{X} \times \mathcal{U} \mapsto \mathcal{R}^n$ is $\mathscr{C}^{D_h}$, i.e., it admits continuous partial derivatives of order $1,\hdots,D_h$ on $\mathcal{X}$. Given $\mathcal{R}^q$ such that $x(t_q) \in \mathcal{R}^q$ and a control signal $u$ that is $\mathscr{C}^{D_u}$ on the interval $[t_q, t_{q+1}]$, interval Taylor-based methods~\cite{berz1998verified,nedialkov1999validated} provide an over-approximation $\mathcal{R}^{q+1}$ of the reachable set at $t_{q+1}$ under the control $v$ as follows:
    \begin{align}
        \mathcal{R}^{q+1} = \mathcal{R}^{q}  + \sum_{d=1}^{D-1} \big( t_{q+1}-t_{q}\big)^d \big( \boldsymbol{h}^{[d]}(\mathcal{R}^{q}, \boldsymbol{v})\big) (t_q) 
                     &+ \big(t_{q+1}-t_{q}\big)^{D} \big( \boldsymbol{h}^{[D]}(\mathcal{P}^{q},\boldsymbol{v})\big)([t_q,t_{q+1}]), \label{eq:taylor-expansion}
    \end{align}
    where $D\leq\min(D_u+1,D_h)$ is the order of the Taylor expansion, $\boldsymbol{h}$ is an interval extension of $h$, $\boldsymbol{v}$ is an interval extension of $v$, $\boldsymbol{h}^{[d]}$ are interval extensions of the Taylor coefficients $h^{[d]}$ defined inductively by
    \begin{align}
        h^{[1]} = h, \: h^{[d+1]} = \frac{1}{d+1}\Big( \frac{\partial h^{[d]}}{\partial x} h + \sum_{l=0}^{d-1} \frac{\partial h^{[d]}}{\partial u ^{(l)}} u^{(l+1)} \Big), \label{eq:taylor-coeff}
    \end{align}
    and the set $\mathcal{P}^q \in \mathbb{IR}^n$ is an a priori rough enclosure of $\{x(t_{q+1}; v^q, x^q) \in \mathcal{X} | x^q \in \mathcal{R}^q\}$ and is a solution of the fixed-point equation
    \begin{align}
        \hspace*{-0.5em} \mathcal{R}^{q} \ + \  [0, t_{q+1}-t_{q}]\ \mathscr{R}(h, \mathcal{P}^{q} \times \boldsymbol{v}([t_q,t_{q+1}])) \ \subseteq \ \mathcal{P}^{q}. \label{eq:rough-enclosure-approx}
    \end{align}
    
    Then, in the setting of our problem, we have that $h(x,u) = f(x) + \sum_{p=1}^d g_p(x) \pU[\alpha^p]$ where $f$ and $g_p$ are unknown functions. Further, with only the Lipschitz assumption on $f$ and $g_p$, we are limited to a Taylor expansion of order $D=2$. By Corollary~\ref{thm:diffinclusion}, $\boldsymbol{h}$ defined in~\eqref{eq:diffinclusion} is a straighforward interval extension of the unknown $h$. Further, since $\mathscr{R}(h, \mathcal{P}^{q} \times \boldsymbol{u}([t_q,t_{q+1}])) \subseteq \boldsymbol{h}(\mathcal{P}^q, \boldsymbol{u}([t_q,t_{q+1}]))$, the set $\mathcal{P}^q$ in Theorem~\ref{thm:one-step-overapprox} is an a priori rough enclosure that satisfies the fixed-point equation~\eqref{eq:rough-enclosure-approx} when $ \boldsymbol{u}([t_q,t_{q+1}]) = u^q$. Thus, we apply the Taylor expansion~\eqref{eq:taylor-expansion} to obtain
    \begin{align}
        \mathcal{R}^{q+1} = \mathcal{R}^q + \Delta t \Big( \boldsymbol{h}(\mathcal{R}^q,u^q) \Big)  + \frac{\Delta t^2}{2} \Big(  \boldsymbol{\frac{\partial h }{\partial x}}  \boldsymbol{h}  \Big) (\mathcal{P}^q, u^q) 
        + \frac{\Delta t^2}{2} \Big( \boldsymbol{\frac{\partial h }{\partial u}} \Big) (\mathcal{P}^q, u^q) \boldsymbol{\dot{u}}([t_q,t_{q+1}]).  \label{eq:taylor-2-order}
    \end{align} 
    Recall that $\boldsymbol{\dot{u}}  = 0$ since the control signal $u = u^q$ is constant on $[t_q,t_{q+1}]$. Furthermore, for all $k,l \in \Nn$, we have 
    \begin{align*}
        \frac{\partial h_k}{\partial x_l}(x,u^q) = \frac{\partial f_k}{\partial x_l}(x) + \sum_{p=1}^d \frac{\partial g_{p,k}}{ \partial x_l}(x) \pU^q[\alpha^p].
    \end{align*}
    Thus, by definition of the upper bounds on the Lipschitz constants of $f_k$ and $g_{p,k}$ we have
    \begin{align*}
        \frac{\partial f_k}{\partial x_l}(x)  \in [-1, 1] w_k \lipf_k \text{ and }
        \frac{\partial g_{p,k}}{ \partial x_l}(x)  \in [-1, 1] w_k \lipg_{p,k}.
    \end{align*}
    Therefore, $\boldsymbol{\frac{\partial h }{\partial x}}(\mathcal{P}^q, u^q) = \mathcal{J}^f + \sum_{p=1}^d \mathcal{J}^{g_p} \pU^q[\alpha^p]$ is an interval extension of the jacobian of $h$ with respect to $x$.
    Finally, merging $\boldsymbol{\frac{\partial h }{\partial x}}$ and $\dot{\boldsymbol{u}} = 0$ into~\eqref{eq:taylor-2-order} provides the over-approximating set~\eqref{eq:nextstate}.

\section*{Proof of Theorem~\ref{thm:suboptimality-gap}}
Let $\mathrm{u}$ and $\mathrm{v}$ be the optimal control values corresponding to the cost $C_j^*$ and $C_j$. For notation simplicity, let $h(w^q, u^q)$ denotes $x(t_{q+1}; u^q, w^q)$. Thus, $w^{q+1}  = h(w^q, u^q)$ and $y^{q+1} = h^\theta(y^q, v^q)$ are completely determined by the control values $u^q, v^q$ and the current state $y^j = w^j = x^j$. We have 
    \begin{align}
        |\hat{C}_j -  C_j^{\star})| &= \begin{cases} \hat{C}_j - \sum_{q=j}^{j+N}c(w^q, u^q, h(w^q,u^q)),& \mathrm{if } \: \hat{C}_j \geq C_j^{\star} \\ C_j^{\star} - \sum_{q=j}^{j+N} c(y^q, v^q, h^\theta(y^q,v^q)),& \mathrm{otherwise} \end{cases} \\
        &\leq \begin{cases} \sum_{q=j}^{j+N}\big(c(y^q, u^q, h^\theta(y^q,u^q)) - c(w^q, u^q, h(w^q,u^q)) \big),\\
        \quad \mathrm{if } \: \hat{C}_j \geq C_j^{\star} \\ 
        \sum_{q=j}^{j+N} \big( c(w^q, v^q, h(w^q,v^q)) - c(y^q, v^q, h^\theta(y^q,v^q)) \big),\\
         \quad \mathrm{otherwise} \end{cases} \\
        &\leq \begin{cases} \sum_{q=j}^{j+N}L_c \big(\|h(w^q,u^q)- h^\theta(y^q,u^q)\|_2 + \|w^q - y^q\|_2\big),\\
        \quad \mathrm{if } \: \hat{C}_j \geq C_j^{\star} \\ 
        \sum_{q=j}^{j+N} L_c \big(\|h(w^q,v^q)- h^\theta(y^q,v^q)\|_2  + \|w^q - y^q\|_2\big),\\
         \quad \mathrm{otherwise} \end{cases} \\
        &\leq L_c \Big( \|\mathrm{wd}(\mathcal{R}_\mathcal{U}^{j+N+1})\|_2  +\sum_{q=j+1}^{j+N} 2\|\mathrm{wd}(\mathcal{R}_\mathcal{U}^{q})\|_2\Big).  
    \end{align}
    The first inequality is obtained by definition of $\hat{C}_j$ and $C_j^\star$ as optimal solutions of the optimal control problem under the different dynamics $h$ and $h^\theta$. That is, for any control other than $u^q$ ($v^q$), the cost returned when rolling out the unknown dynamics given by $h$ ($h^\theta$) is suboptimal. The second inequality uses the definition of the Lipschitz constant of $c$. Finally, in the last inequality, we use the fact that $h(w^q,u^q), h(w^q,v^q) \in \mathcal{R}_\mathcal{U}^{q+1}$ and $h^\theta(y^q, v^q),h^\theta(y^q, u^q) \in \mathcal{R}_\mathcal{U}^{q+1}$ to conclude.

\section*{Incorporating More Side Information}
We recall that general constraints on the states or its derivatives, in the form of nonlinear mathematical (in)equalities, can be used as side information via the interval contractor programming framework. In the following, we provide others examples of side information that can be incorporated in the proposed learning algorithm.

\begin{sideinfo}[\textsc{Decoupling among states}]\label{side:state-dependency}
    The qualitative knowledge that some components of the vector field $\dot{x}$ do not depend on some components of the state $x$. In other words, the subset of states for which some components of $f$ and $g_p$ are independent is known. 
    
    For example, if the state $x_l(t)$ does not directly affect $\dot{x}_k(t)$ for some $l,k \in \Nn$ under any control signal in $ \mathcal{U}$, we can obtain a tighter over-approximation of the reachable set by setting to zero the intervals $\mathcal{J}^f_{k,l}$ and $\mathcal{J}^{g_p}_{k, l}$ for all $p \in \Nd$.
\end{sideinfo}

\begin{sideinfo}[\textsc{Gradient bounds}]\label{side:gradB}
    We are given bounds on the gradient of some components of $f$ and $g_p$. Such side information may include the monotonicity of $f$ or $g_p$.
    
    These bounds can be used to provide tight interval extensions $\mathcal{J}^f$ and $\mathcal{J}^{g_p}$. For example, if the function $f_k$ is known to be non-decreasing with respect to the variable $x_l$ on a set $\mathcal{A} \subseteq \mathcal{X}$, then we obtain a tighter $\mathcal{R}^{q+1}$ by the update $\mathcal{J}^f_{k,l} \gets \mathcal{J}^f_{k,l} \cap \mathbb{R}_+$ if $\mathcal{P}^q \subseteq \mathcal{A}$.
\end{sideinfo} 

\begin{sideinfo}[\textsc{Vector field bounds}]\label{side:vFieldB}
    We are given the sets $\mathcal{R}^{f_\mathcal{A}} \in \mathbb{IR}^n$ and $\mathcal{R}^{G_{\mathcal{A}}} \in \mathbb{IR}^{n \times m}$ as supersets of the range of $f$ and $g_p$, respectively, over a given set $\mathcal{A} \subseteq \mathcal{X}$.

	Given a set $\mathcal{S} \subseteq \mathcal{A}$, tight extensions of $\boldsymbol{f}$ and $\boldsymbol{g}_p$ over $\mathcal{S}$ can be obtained by the update $\boldsymbol{f}(\mathcal{S}) \gets \boldsymbol{f}(\mathcal{S}) \cap \mathcal{R}^{f_\mathcal{A}}$ and $\boldsymbol{g}_p(\mathcal{S}) \gets \boldsymbol{g}_p(\mathcal{S}) \cap \mathcal{R}^{G_\mathcal{A}}$. The tight extensions can be directly used in Corollary~\ref{thm:diffinclusion}, the refinement algorithm, and the initialization of the construction algorithm.
\end{sideinfo} 

\section*{Sequential Convex Programming Algorithm}

\begin{algorithm}[!hbt]
    \caption{Sequential convex programming with trust region to find an approximate solution to the $N$-step optimal control problem.}\label{alg:scp}
    \textbf{Input: } Dataset $\traj_j$, current system's state $x^j$, initial trust region $r^1 > 0$, penalty weight $\lambda > 0$, trust region parameters $0<\rho_0<\rho_1<1$, $\alpha > 1$, and optimality tolerance $\epsilon_{\mathrm{tol}} > 0$. \\
    \textbf{Output: } $\mathrm{u}, \mathrm{x}$ %\COMMENT{Solution to the N-step control problem} 
    \begin{algorithmic}[1]    
        \STATE Initialize $\mathrm{x}^1 \in \mathcal{X}^N, \mathrm{u}^1 \in \mathcal{U}^N$, $s \gets 1$ \hfill \COMMENT{// The starting point does not need to be feasible}
        %\State $s \gets 1$
        \WHILE{$\mathrm{True}$}
            \STATE Solve linearized problem~\eqref{eq:convexsub} at $\mathrm{x}^s, \mathrm{u}^s, r^s$ to obtain $\Delta \mathrm{x}^{s+1}, \Delta \mathrm{u}^{s+1}$ 
            \STATE $\Delta J^s \gets J(\mathrm{x}^s, \mathrm{u}^s) - J(\mathrm{x}^s + \Delta \mathrm{x}^{s+1}, \mathrm{u}^s + \Delta \mathrm{u}^{s+1})$ \label{algo:actual-change}\hfill \COMMENT{// Variation of the realized cost}
            \STATE $\Delta L^s \gets J(\mathrm{x}^s, \mathrm{u}^s) - L^s(\Delta \mathrm{x}^{s+1}, \Delta \mathrm{u}^{s+1})$ \hfill \COMMENT{// Variation of the linearized cost}
            \IF{$|\Delta J^s| \leq \epsilon_{\mathrm{tol}}$} 
                \STATE \RETURN $\mathrm{x}^s, \mathrm{u}^s$ \hfill \COMMENT{// Found solution}
            \ENDIF
            \STATE $\rho^s \gets \Delta J^s / \Delta L^s$ \hfill \COMMENT{// Encode the quality of linearization}
            \IF{$\rho^s < \rho_0$}
                \STATE $r^s \gets r^s / \alpha$ \hfill \COMMENT{// Contract trust region}
            \ELSE 
                \STATE $s \gets s + 1$ \hfill \COMMENT{// Update estimate}
                \STATE $\alpha^s \gets r^s / \alpha \:  \boldsymbol{\mathrm{if}} \: \rho^s \leq \rho_1 \: \boldsymbol{\mathrm{else}} \: r^s \alpha$ \hfill \COMMENT{// Contract or expand trust region}
            \ENDIF
        \ENDWHILE
    \STATE \RETURN $\mathrm{x}^s, \mathrm{u}^s$
  \end{algorithmic}
\end{algorithm}

Algorithm~\ref{alg:scp} summarizes the trust-region-based sequential convex optimization scheme to compute approximate (possibly local) solutions to the $N$-step optimal control problem~\eqref{eq:n-step-optimal-control}. Specifically, the quality of the linear approximation can be understood by inspecting the ratio $\rho^s$. The ratio $\rho^s$ compares the realized reduction $\Delta J^s$ to the predicted cost $\Delta L^s$. When $\rho^s \leq \rho_0$ with $\rho_0$ sufficiently close to $0$, the linearization is considered inaccurate. Then, we contract the trust region $r^s$ and restart the iteration. If not, the solutions $\Delta \mathrm{x}^{s+1}$ and $\Delta \mathrm{u}^{s+1}$ are considered acceptable. Then, we move to the next iteration and contract or expand the trust region depending on if $\rho^k$ is below or above the threshold $\rho_1$ typically chosen to be close to $1$.

\section*{Implementation Details}

All the experiments in this paper were performed on a computer with an Intel Core $i9$-$9900$ CPU $3.1$GHz $\times 16$ processors and $31.2$ Gb of RAM. We use Gurobi $9.0$~\cite{gurobi} to solve each subproblem of the sequential convex optimization~\eqref{eq:convexsub}, and we used the control tasks of DeepMind Control Suite~\cite{tassa2018deepmind} for comparison with RL algorithms.

\paragraph{Code.} All the implementations are written and tested in Python $3.8$, and we will release the full code at \url{https://github.com/wuwushrek/datacontrolreach.git}. We emphasize that this code is still under development and can be significantly improved both on its organization and efficiency.

\paragraph{Comparisons with \sac{} and \dpg{}.} We utilized the implementation of \dpg{} from~\cite{hoffman2020acme} and \sac{} from~\cite{pytorch_sac} for a comparison with the proposed data-driven control algorithm. The default hyper-parameters provided by the implementations of \dpg{} and \sac{} were used to train the agents in the Reacher, Swimmer, and Cheetah environment. Such hyper-parameters have been empirically demonstrated to provide high performance in control tasks of MuJoCo. We recall that \dpg{} and \sac{} were pre-trained using \emph{ten} million of iterations with the environment. Then, the testing phase was done for $100$ episodes. Our algorithm only learns from the data obtained during the $1000$ time steps of each episode.

\paragraph{Experiments in MuJoCo.} We recall that the Reacher environment model has been slightly modified to accommodate our learning algorithm. Specifically, our algorithm works well with smaller time steps and we modified the original Reacher model to accommodate to that. We provide in the code the new Reacher model description to be used inside the DeepMind Control Suite framework. The others environments were not modified as the time steps were already small enough for our algorithm to succeed in learning the control tasks.

The Lipschitz bounds were estimated using trajectories generated by an excitation-based control of the system. We provide a code to compute under-estimation of the Lipschitz bounds for any MuJoCo environment. In addition, the nonconvex cost functions were linearized as detailed in this paper using finite differentiation and the well-documented API provided by MuJoCo.

\section*{Experiments on the F-16 Aircraft Simulator}
We provide in this section additional details on the F-16 aircraft simulator since the simulator is not publicly available yet.
\begin{figure}[!hbt]
    \centering
    \frame{\includegraphics[scale=0.103]{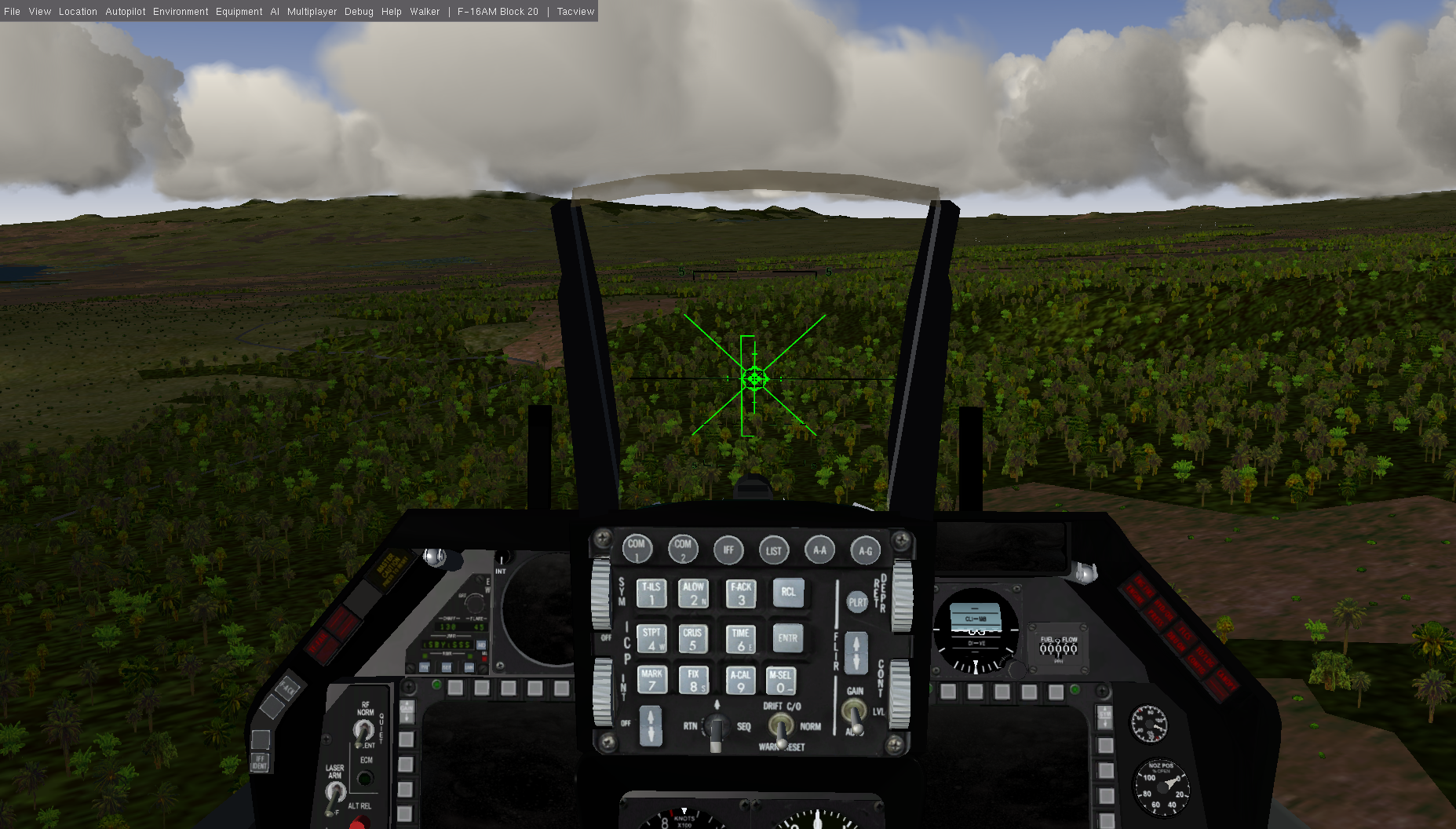}}
    \frame{\includegraphics[scale=0.1]{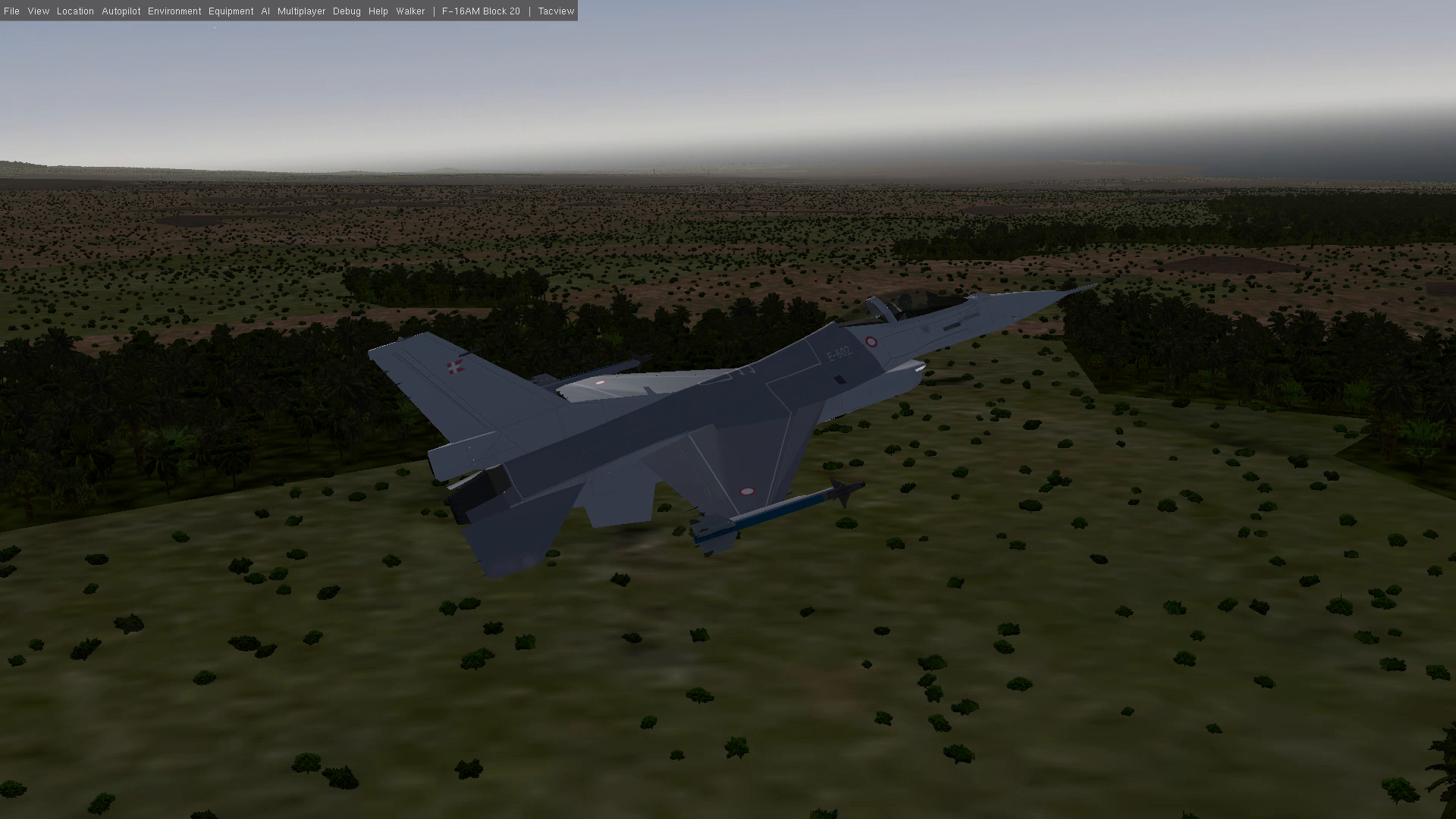}}
    \caption{Screenshots of the inside and outside views of the F-16 aircraft in FlightGear simulation.}
    \label{fig:f16}
\end{figure}

The F-16 aircraft's flight control system~\cite{heidlauf2018verification} is described by a hierarchical feedback control loop consisting of an autopilot and a low-level controller. The autopilot performs \textit{higher-level} maneuvers, such as ground collision avoidance, waypoint tracking, and more. In contrast, the low-level control tracks the references from the autopilot and maintains stability by actuating the flight control surfaces appropriately. The control system uses a closed-loop feedback control to actuate the flight control surfaces, including the thrust, the ailerons, elevators, and rudders, in order to meet the desired flight objectives.

Inside the simulator, the underlying nonlinear dynamics, containing $13$-states and $4$-control inputs, capture the ($6$-DOF) movement of an aircraft through the standard aerodynamic equations. The dynamics describe the evolution of the system's states, namely velocity $v_t$, angle of attack $\alpha$, sideslip $\beta$, altitude $h$, attitude angles: roll $\phi$, pitch $\theta$, yaw $\psi$, and their corresponding rates $p$, $q$, $r$, engine $power$ and two more states for translation along north and east. The plant model is built on several \emph{linearly interpolated lookup tables} that incorporate wind tunnel data describing the engine model, the various coefficients including damping, force and moment coefficients, and the moments due to the control surfaces. As a consequence, with known look-up tables, the resulting dynamics have \emph{polynomial dependency} in the control input.

\paragraph{Side Information and F-16 Dynamics.} Our algorithm considers as side information the following knowledge of the rigid-body dynamics of a $6$-DOF~\cite{stevens2015aircraft}
\begin{equation} \label{eq:gaim-dynamics}
\begin{aligned}
\dot{u} &= rv - qw - g \sin \theta + \frac{F_u}{m}, \\
\dot{v} &= -ru + pw + g \sin \phi \cos \theta + \frac{F_v}{m}, \\
\dot{w} &= qu - pv + g \cos \phi \cos \theta + \frac{F_w}{m}, \\
\dot{v}_t &= \frac{u \dot{u} + v \dot{v} + w \dot{w}}{v_t},\\
\dot{\alpha} &= \frac{u \dot{w} - w \dot{u} }{u^2 + w^2},\\
\dot{\beta} &= \frac{(v_t \dot{v} - v \dot{v}_t) \cos \beta }{u^2 + w^2},\\
\dot{p} &= \frac{J_y -J_z}{J_x} qr + \frac{M_p}{J_x},\\
\dot{q} &=  \frac{J_z -J_x}{J_y} pr + \frac{M_q}{J_y}, \\
\dot{r} &=  \frac{J_x -J_y}{J_z} pq + \frac{M_r}{J_z},\\
\dot{\phi} &= p + \tan \theta (q \sin \phi + r \sin \phi),\\
\dot{\theta} &= q \cos \phi - r \sin \phi, \\
\dot{\psi} &= \frac{q \sin \phi + r \cos \phi}{\cos \theta},
\end{aligned}
\end{equation}
where $x = [v_t, \alpha, \beta, p, q, r, \phi, \theta,\psi, \mathrm{power}, h, p_n, p_e]$ is the full state vector of the aircraft, the intermediary variables $u = v_t \cos \alpha \cos \beta$, $v = v_t \sin \beta$, and $w = v_t \sin \alpha \cos \beta$ represent respectively the axial, lateral, and vertical velocities in the body frame, $v_t$ is the truth velocity, $\alpha$ is the angle of attack, $\beta$ is sideslip, $p$ is the pitch rate, $q$ is the roll rate, $r$ is the yaw rate, $\phi$ is the roll angle, $\theta$ is the pitch angle, $\psi$ is the yaw angle, $h$ is the altitude, $\mathrm{power}$ is the resulting power when applying thrust, $p_n$ is the position on the north axis, and $p_e$ is the position on the east axis. The aerodynamics forces and moments are given by $F_u, F_v, F_w$ and $M_p, M_q, M_r$, respectively. \emph{Such possibly time-varying forces and moments depend on the wing and control surfaces, the states, and the control inputs of the aircraft}. That is, we have the dependencies $F_u(x, u)$, $F_v(x, u)$, $F_w(x, u)$, $M_p(x, u)$, $M_q(x,u)$, $M_r(x, u)$, where $u = [\mathrm{thrust}, \delta_a, \delta_e, \delta_r]$ is the vector of control inputs. Here, $\delta_a, \delta_e$, and $\delta_r$ are the aileron, elevator, and rudder control inputs, respectively, $m$ is the mass of the aircraft, $J_x, J_y$, and $J_z$ are the inertia moments, and $g$ is the gravity constant.

Specifically, even though the form of the dynamics above is known, \emph{we consider that the $\mathrm{power}$ dynamics, the forces and moments, typically approximated via lookup tables and experiments, are unknown functions of the states and control inputs. In other words, the effect of the control inputs on the aircraft are unknown but we still want to retain some degree of control using streaming data from a single trial.} Such unknown functions are nonlinear in the states and polynomial in the control inputs.

\begin{figure}[!hbt]
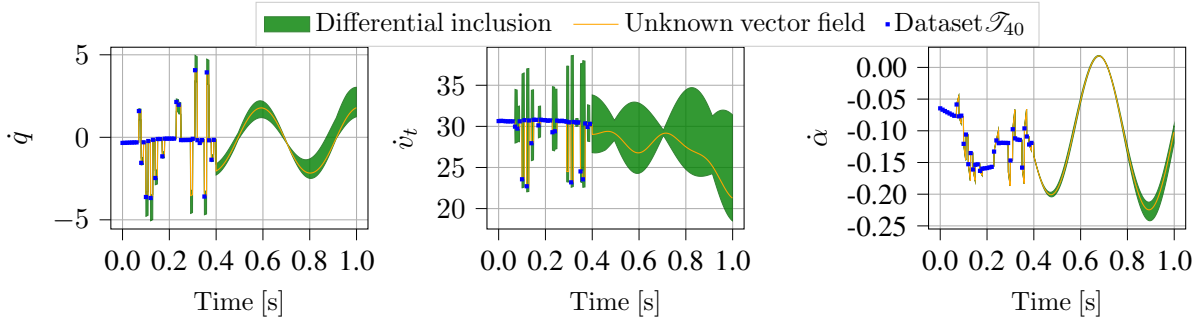

    \centering
    \hspace*{-1.5cm}
    \input{texf16/fig6old}
    \hspace*{-2.0cm}
    % \hfill
    \input{texf16/fig0old}
    \hspace*{-3.5cm}
    \input{texf16/fig1old}
    % \begin{subfigure}{0.3\textwidth}
    %     \hspace*{-1.5cm}
    %     \centering
    %     \input{texf16/fig6old}
    %     \phantomsubcaption
    % \end{subfigure}
    % \hfill 
    % \begin{subfigure}{0.3\textwidth}
    %     \vspace*{-0.05cm} \hspace*{-2.80cm}
    %     \centering
    %     \input{texf16/fig0old}
    %     \phantomsubcaption
    % \end{subfigure}
    % \hfill
    % \begin{subfigure}{0.3\textwidth}
    %     \vspace*{0.50cm} \hspace*{-0.70cm}
    %     \centering
    %     \input{texf16/fig1old}
    %     \phantomsubcaption
    % \end{subfigure}
    \caption{From left to right, we show how the constructed data-driven differential inclusion can be indeed tight on the F-16 aircraft example.}
    \label{fig:diff-illu}
    % \vspace*{-2.5em}
\end{figure}

\paragraph{Data-Driven Differential Inclusion.} We empirically demonstrate in Figure~\ref{fig:diff-illu} that the data-driven differential inclusion constructed from streaming data from the ongoing trajectory and Lipschitz bounds is indeed tight.

\paragraph{Ground Collision Scenario.} In the ground collision avoidance scenario, we initialize the simulator such that the plane is diving nose down towards the ground with an extremely high downward pitch angle. The autopilot uses a PID law to compute the references on the system's states that our algorithm must track to avoid the crash. The initial condition considered is given by $\theta=-85 \pi / 180, \: v_t = 540,  \: h = 3600, \phi = \pi/4, \: \psi = - \pi /4, \beta = 0, \alpha = 2.5 \pi / 180, \: p = q = r = 0$.

At each time step, the autopilot computes and adjusts the state setpoints required to avoid the ground collision. Thus, the cost function that our algorithm is trying to optimize is given by: $\mathrm{cost}(x,u) = \|x - x_{\mathrm{target}}\|_2$, where $x_{\mathrm{target}}$ is the target provided by the autopilot. We compare the highly-tuned LQR controller of the simulator to our algorithm trying to avoid ground collision from streaming data obtained while diving towards the ground.

% % \begin{figure}[!hbt]
% %     \centering
% % 	\input{simu_plot_supp}
% % % 	\vspace*{-1em}
% % 	\caption{The algorithm developed in this paper successfully enables the F-16 to avoid the collision with the ground while the embedded, pre-tuned \texttt{LQR} controller fails to avoid the crash due to the low altitude and extremely high downward pitch angle. In addition, the algorithm can be applied in real time as its computation time is less than the \emph{control time step} $\Delta t = 0.01$s enforced by the simulator.}
% % \end{figure}